\documentclass[useAMS, usenatbib]{mn2e}  
\usepackage{graphicx}
\usepackage{bm}
\usepackage[T1]{fontenc}
\usepackage[varg]{txfonts}
\usepackage{epstopdf}

\usepackage{color}

\newcommand{\be}{\begin{equation}}
\newcommand{\ee}{\end{equation}}

\title[Environment-based selection effects of Planck clusters]{Environment-based selection effects of Planck clusters}
\author[R.\, Kosyra et al.]{\parbox{0.9\textwidth}{R.\, Kosyra$^{1,2}$, 
       	D.\, Gruen$^{1,2}$,
 	S.\, Seitz$^{1,2}$,	
 	A.\, Mana$^{1,2}$,
        E.\, Rozo$^{3,4}$,
        E.\, Rykoff$^{3}$, 
        A.\, Sanchez$^{2}$ and
	R.\, Bender$^{1,2}$
	 \small \thanks{E-mail: kosyra@usm.uni-muenchen.de} \\
	$^{1}$Universit{\"a}ts-Sternwarte M{\"u}nchen,  Ludwig-Maximilians-Universit{\"a}t M{\"u}nchen, Scheinerstra{\ss}e~1, D-81679~M{\"u}nchen, Germany\\
	$^{2}$Max Planck Institut f\"ur Extraterrestrische Physik, Giessenbachstr., D-85748 Garching, Germany\\
	$^{3}$SLAC National Accelerator Laboratory, Menlo Park, CA 94025, U.S.A.\\
	$^{4}$University of Arizona, Department of Physics, 1118 E. Fourth St., Tucson, AZ 85721, U.S.A.}}
	
\begin{document}



\maketitle

\begin{abstract}
We investigate whether the large scale structure environment of galaxy clusters imprints a selection bias
on Sunyaev Zel'dovich (SZ) catalogs. Such a selection effect might be caused
by line of sight (LoS) structures that add to the SZ signal or contain point sources that disturb the signal extraction in
the SZ survey. We use the \textit{Planck} PSZ1 union catalog \citep{2013arXiv1303.5089P} in the SDSS region as our sample
of SZ selected clusters. We calculate the angular two-point correlation 
function (2pcf) for physically correlated, foreground and background structure in the \textit{RedMaPPer} SDSS DR8 catalog with respect to each cluster.
We compare our results with an optically selected comparison cluster sample and with
theoretical predictions. In contrast to the hypothesis of no environment-based selection, we find a mean 2pcf for background structures of
$-0.049$ on scales of $\lesssim 40'$, significantly non-zero at $\sim\!\!4 \sigma$, which means that \textit{Planck} clusters are more likely to be detected in regions of
low background density.
We hypothesize this effect arises either from background estimation in the SZ survey or from radio sources in the background.
We estimate the defect in
SZ signal caused by this effect to be negligibly small, of the order of $\sim 10^{-4}$ of the signal of a typical
\textit{Planck} detection. Analogously, there are no implications on X-ray mass measurements. 
However, the environmental dependence has important consequences for weak lensing follow up of \textit{Planck} 
galaxy clusters: we predict that projection effects account for half of the mass contained within a 15' radius of \textit{Planck} galaxy clusters.
We did not detect a background underdensity of CMASS LRGs, which also leaves a spatially varying redshift dependence of the \textit{Planck}
SZ selection function as a possible cause for our findings.
\end{abstract}


\section{Introduction}

Clusters of galaxies play a major role in astrophysics and cosmology, as they can be used to put constraints on the dark matter content of the universe.
Furthermore galaxy clusters are particularly sensitive to the interplay of dark matter and dark energy. They are 
cosmological probes that could potentially help to distinguish between dark energy and modified gravity
explanations for the accelerating expansion of the universe \citep[for a review, see ][]{2011ARA&A..49..409A,2011ASL.....4..204B,2013arXiv1309.5380W}.

A variety of different methods for cluster detection and mass measurement exists. Gravitational lensing probes
the dark and luminous matter distribution of a cluster by measuring the distortion of background galaxies 
\citep[weak lensing, for example in][]{2001ApJ...548L...5H,2013MNRAS.432.1455G,2014MNRAS.442.1507G}, or by detecting multiple images 
of single background galaxies close to the LoS of the cluster core 
\citep[strong lensing, for example in][]{2012ApJ...749...97Z,2013ApJ...774..124E,2014MNRAS.438.1417M}. 
The most widespread method for optical cluster detection, the so-called \textit{red sequence method}
 \citep{2005ApJS..157....1G,2007ApJ...660..239K,2014ApJ...785..104R} is based on spatial overdensities of red galaxies.
Further methods include the observation of the X-ray Bremsstrahlung emission by the hot gas in the intra-cluster medium  
\citep[ICM,][]{2011A&A...534A.109P,2009ApJ...692.1060V,2010MNRAS.406.1759M}
and the observation of inverse Compton scattering of the cosmic microwave background (CMB)
photons by the ICM, which is known as the SZ effect \citep[]{1972CoASP...4..173S}.
The latter describes the distortion of the CMB spectrum along the LoS through clusters and groups.
The amplitude of the SZ effect is proportional to the dimensionless Compton parameter $y$,  defined as the integral over the thermal
electron pressure along the LoS:
 \begin{equation}
  y=\frac{\sigma_T}{m_e c^2} \int P dl \, , 
 \end{equation}
  while the integral over a solid angle yields the SZ observable $Y$:
  \begin{equation}
   D_A^2 Y= D_A^2 \int y d\Omega =  \frac{\sigma_T}{m_e c^2} \int P dV \, ,
  \end{equation}

 where $\sigma_T$ is the Thomson cross section, $m_e c^2$ the rest energy of the electrons and $D_A$ the
 angular diameter distance.

 All of these methods may have selection effects induced by structures along the LoS. Lensing, for example can yield biased mass
 estimates when there are groups along the LoS which contribute to the shear signal \citep[e.g.][]{2012MNRAS.420.1384S}. 
 The X-ray signal of LoS structures can stack, resulting in
 a biased mass estimate. The same is true for the SZ effect, however more severely as the SZ signal is proportional to the gas density $\rho$, while the 
X-ray flux is proportional to $\rho^2$, making the effect of LoS structure on SZ signals much larger at larger angular separation.
Due to this reason we will investigate whether SZ selected clusters are possibly biased by structures along the LoS,
either by physically uncorrelated foreground or background structures, or by correlated structures at the same redshift as the cluster itself.

Several effects could potentially contribute to a selection bias. The blending of the SZ signal of the detected cluster with groups along the LoS
could bias the SZ estimate high and cause clusters along overdense lines of sight to be more likely detected.
If, on the contrary, unresolved groups in the vicinity of clusters increase the background level, this 
could lead to a lower detection probability as the signal from the cluster is partly suppressed by the wrong background estimate.
Furthermore, if the background of a cluster is contaminated with radio-loud galaxies,
this could raise the noise such that clusters with a weak SZ signal are not detected.

In this paper we address this question by analyzing the projected group environment of SZ-selected clusters from the \textit{Planck} PSZ1 union
catalog \citep{2013arXiv1303.5089P} and test for group overdensities or underdensities along the LoS 
in the foreground, background and at the redshift of the clusters. The group sample is taken from the \textit{RedMaPPer} red-sequence catalog
based on SDSS DR8 photometry \citep{2014ApJ...785..104R,2014ApJ...783...80R,2014arXiv1401.7716R}.
We compute the angular two-point correlation function (2pcf) of galaxy clusters and groups for different subsamples of our catalogs 
(correlated, foreground and background structures) to quantify correlated and physically uncorrelated
group overdensities and underdensities. We compare these results to the 2pcf obtained for an independent cluster sample with similar redshift and richness
distribution, drawn as a subsample of the \textit{RedMaPPer} SDSS DR8 catalog, and to theoretically predicted values.

\subsection{Motivation}

We briefly discuss several possible effects that could cause a selection bias.
The filter function that is used for the \textit{Planck} cluster detection might estimate a too large background
value if there are groups surrounding the cluster that contribute to the signal, which could lead to a decreased detection probability
in crowded fields as the subtracted background estimate is too large. On the other hand, the clusters are detected by combining six frequency
bands with different filter sizes, so it is rather unlikely that this still causes problems
when detecting clusters based on the differential signal.

Another possible origin of a selection effect might be radio-loud galaxies in the background. \citet{2010MNRAS.407.1078D} state that radio-loud
active galactic nuclei (RLAGNs) are predominantly found in dense environments compared to radio quiet galaxies and regular red luminous galaxies (LRGs)
at redshifts $0.4<z<0.8$. They conclude that this clustering effect is stronger for more massive RLAGNs. In \citet{1989MNRAS.240..129Y} the
clustering effect of RLAGNs at $z\approx 0.2$ is compared to the one at $z \approx 0.5$, with the result that the latter objects are found in environments
three times denser on average. They also state that more powerful RLAGNs are found in denser environments than less powerful ones. Based on these findings
we hypothesize that a high background group density entails a higher probability of containing radio sources and thus increases the noise along the 
line of sight, potentially leading to a lower SZ detection probability for clusters in dense background environments.

This paper is structured as follows.
In section \ref{sect2}, we describe the \textit{Planck} PSZ1 union catalog and the \textit{RedMaPPer} SDDS DR8 group catalog, as well as our matching
of these two. In section \ref{sect3}, we briefly discuss two-point correlation functions. Furthermore
we describe our method of generating random points for the \textit{Planck} catalog and the procedure
of defining the cluster comparison sample out of the \textit{RedMaPPer} catalog. We also include the description of our theoretical prediction of the
2pcf. In section \ref{sect4}, we present our results, give a 
detailed description of our error estimation and generalized $\chi^2$ analysis and we estimate the implications of the measured effect
on SZ and lensing analyses of \textit{Planck} clusters. We conclude in section \ref{sect5}.

\section{Data}
\label{sect2}

\subsection{The Planck PSZ1 catalog}

The \textit{Planck} PSZ1 union catalog is a cluster catalog, covering the whole sky based on SZ detections using the first 15.5 months of \textit{Planck}
survey observations. It contains a total of 1227 clusters, 861 of which are confirmed while the remaining 366 are cluster candidates
\citep{2013arXiv1303.5089P}.
The \textit{Planck} satellite features a low frequency and a high frequency instrument, the former covers the bands at 30, 44 and 70 GHz
\citep{2013arXiv1303.5063P} while the latter operates at frequencies of 100, 143, 217, 353, 545 and 857 GHz \citep{2013arXiv1303.5067P}
with angular resolutions between 9.53' and 4.42' FWHM, for a total of nine detection bands.
The channel maps of the six highest frequency bands (100 to 857 GHz) were used to build the SZ-detection catalog, in order to avoid problems
caused by strong radio point sources in cluster centers, which typically have steep spectra and thus do not appear in the high frequency
bands \citep{2013arXiv1303.5089P}. 

The generalized NFW \citep{ 1997ApJ...490..493N} profile from \citet{2010A&A...517A..92A} was adopted for the cluster detection.

Three detection algorithms were used to create the cluster catalog, two realizations of the \textit{Matched Multi-filter (MMF)} method
\citep{2002MNRAS.336.1057H,2006A&A...459..341M} and  \citep[\textit{Powell Snakes (PwS),}][]{2009MNRAS.393..681C,2012MNRAS.427.1384C}.

The \textit{MMF} method detects clusters by using a linear combination of maps and a spatial filtering to suppress foregrounds and noise.
The two implementations (\textit{MMF1} and \textit{MMF3}) split the whole sky in 640 patches of size 14.66 $\times$ 14.66 square degrees 
covering 3.33 times the area of the sky (\textit{MMF1}), and in 504 patches of size 10 $\times$ 10 square degrees covering 1.22 times
the area of the sky (\textit{MMF3}).
The \textit{MMF3} algorithm is run in two iterations: the second is centered on the positions of
the candidates from the first one, rejecting all candidates that fall below the signal-to-noise (S/N) threshold. The matched multi-frequency
filter optimally combines the six frequencies of each patch and the resulting sub-catalogs for all patches are finally merged together to 
a single SZ-catalog per method, selecting the candidate with the highest S/N ratio. For estimating the candidate size, the patches are filtered over the range
of potential scales, selecting the scale with the highest S/N of the current candidate.
Finally, the SZ-signal is estimated by running \textit{MMF} with fixed cluster size and position.

\textit{Powell Snakes} is a Bayesian multi-frequency detection algorithm, optimized to find compact objects in a diffuse background. 
After cluster detection, \textit{PwS} merges all intermediate sub-catalogs.
The cross-channel covariance matrix is calculated directly from the pixel data, which is done in an iterative way to minimize the contamination of the
background by the SZ signal itself. In each iteration step, all detections in the same patch with higher S/N than the current target are subtracted from
the data before re-estimating the covariance matrix. This so-called ``native'' mode of background subtraction produces S/N values 20\% higher than those of the
\textit{MMF} method. In order to emulate the estimation of the background noise cross-power spectrum of the \textit{MMF} method,
\textit{PwS} is run in ``compatibility'' mode, skipping the re-estimation step.

Each of the three detection algorithms creates a catalog of SZ sources with an S/N ratio $\ge$ 4.5.
Obvious false detections are removed from each of the three individual catalogs \citep{2013arXiv1303.5089P}.

The union catalog contains all sources that have been detected by at least two algorithms with S/N $\ge$ 4.5 within a distance of 5$'$, 
fixing the position of the \textit{MMF3} detection or, in case of no \textit{MMF3} detection, keeping the position of the \textit{PwS} detection.

\subsection{The \textit{RedMaPPer} SDSS DR8 catalog}

The Red Sequence Matched-filter Probabilistic Percolation (\textit{RedMaPPer}) algorithm \citep[]{2014ApJ...785..104R} is a red-sequence cluster finder
based on the optimized richness estimator $\lambda$ \citep[]{Rykoff2012}.  It has excellent photo-z performance and $\lambda$ 
has been designed to be a low-scatter mass proxy \citep[]{2014ApJ...783...80R,2014arXiv1401.7716R}. The algorithm is divided into two stages.
The first is a calibration stage where the red sequence model is derived directly from the data by relying on spectroscopic
galaxies in galaxy clusters: given an initial model of the red-sequence, \textit{RedMaPPer} selects cluster member galaxies, uses these to derive
a new red-sequence model, and then iterates the whole procedure until convergence is achieved, as which point the red-sequence
model is adequately calibrated.  The second is the cluster-finding stage, where \textit{RedMaPPer} utilizes the red-sequence model to search for clusters
around every galaxy in the SDSS.  This work uses the updated version (v5.10) of the original \textit{RedMaPPer} catalog of \citet{2014ApJ...785..104R}
presented in \citet{2014arXiv1401.7716R}.

\subsection{Matching of the Planck and \textit{RedMaPPer} catalogs}

In order to match the \textit{Planck} PSZ1 union catalog to the \textit{RedMaPPer} SDSS DR8 cluster catalog we used an algorithm similar to
the one described in \citet{2014arXiv1401.7716R}. We find all matches in the \textit{RedMaPPer}
catalog in a radius of 10$'$ around each \textit{Planck} cluster. In the case of multiple
matches we define the best match as the \textit{RedMaPPer} system with the highest richness. We flag all matches with a redshift difference between the
\textit{Planck} and \textit{RedMaPPer} redshift of more than 3$\sigma$, where $\sigma$ corresponds to the redshift error given in the \textit{RedMaPPer} catalog.
This gives us a total of 290 matched clusters.
 
\subsubsection*{Outlier rejection}

We reject all matches that are obvious SZ-projections (5 cases), as identified by \citet{2014arXiv1401.7716R}. All clusters that have been flagged as 3$\sigma$ 
redshift outliers
are cross-matched with the \citet{2014arXiv1401.7716R} table of redshift outliers, and in the case of an incorrect \textit{Planck} redshift and a correct
\textit{RedMaPPer} redshift, we accept the cluster using the \textit{RedMaPPer} redshift and vice versa. Furthermore we reject clusters with bad z-matching when a visual
inspection identified them clearly as a mismatch (one case only), and we reject clusters that are outliers in the mass-$Y_{SZ}$-plane (according to 
\citealt{2014arXiv1401.7716R}) due to a low \textit{RedMaPPer} richness (one case only). After rejecting all outliers the final matched catalog includes 265 clusters.

%

 \begin{figure*}
  \centering
  \resizebox{\hsize}{!}{\includegraphics{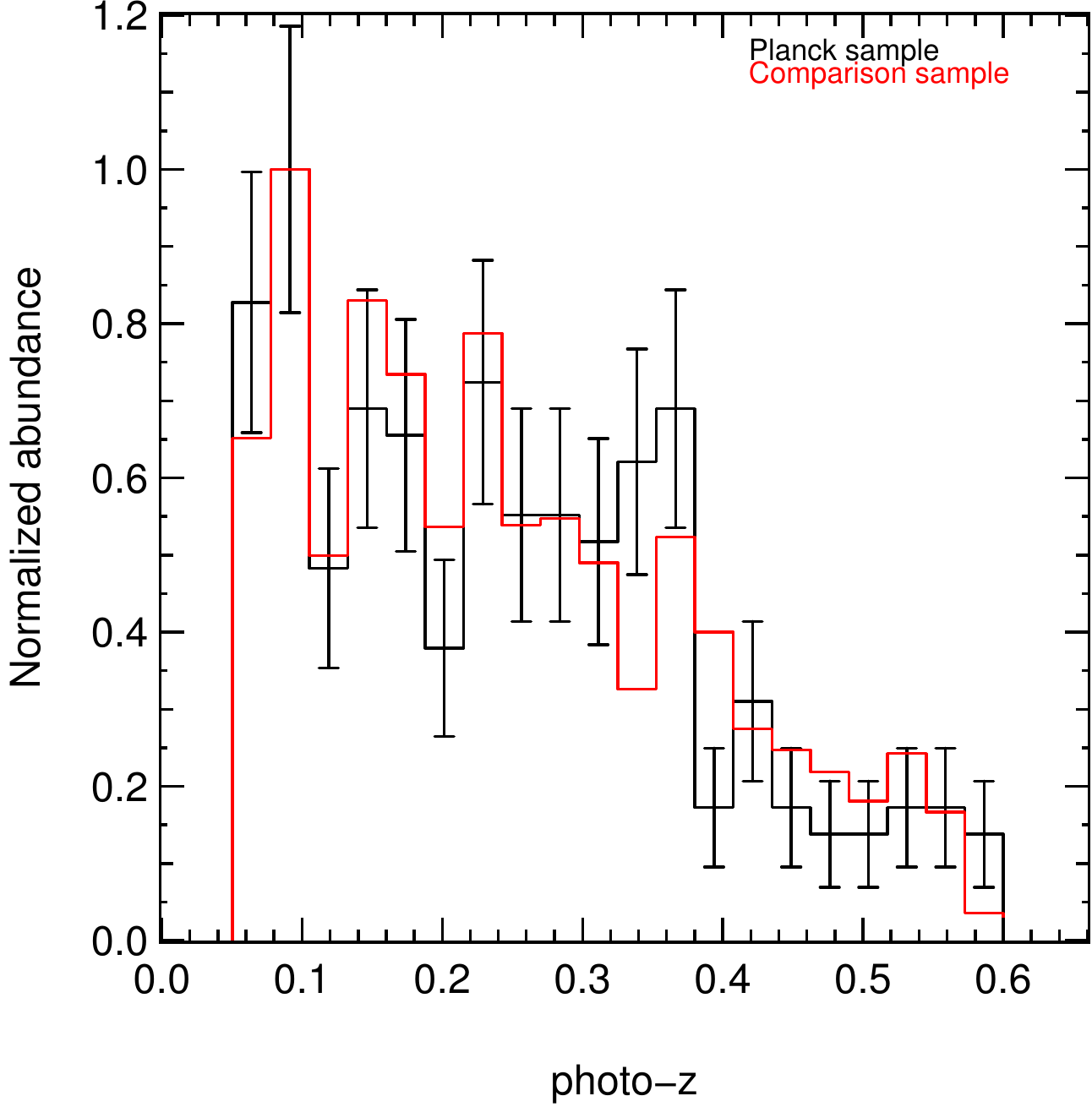}
  \hspace*{20mm}
  \includegraphics{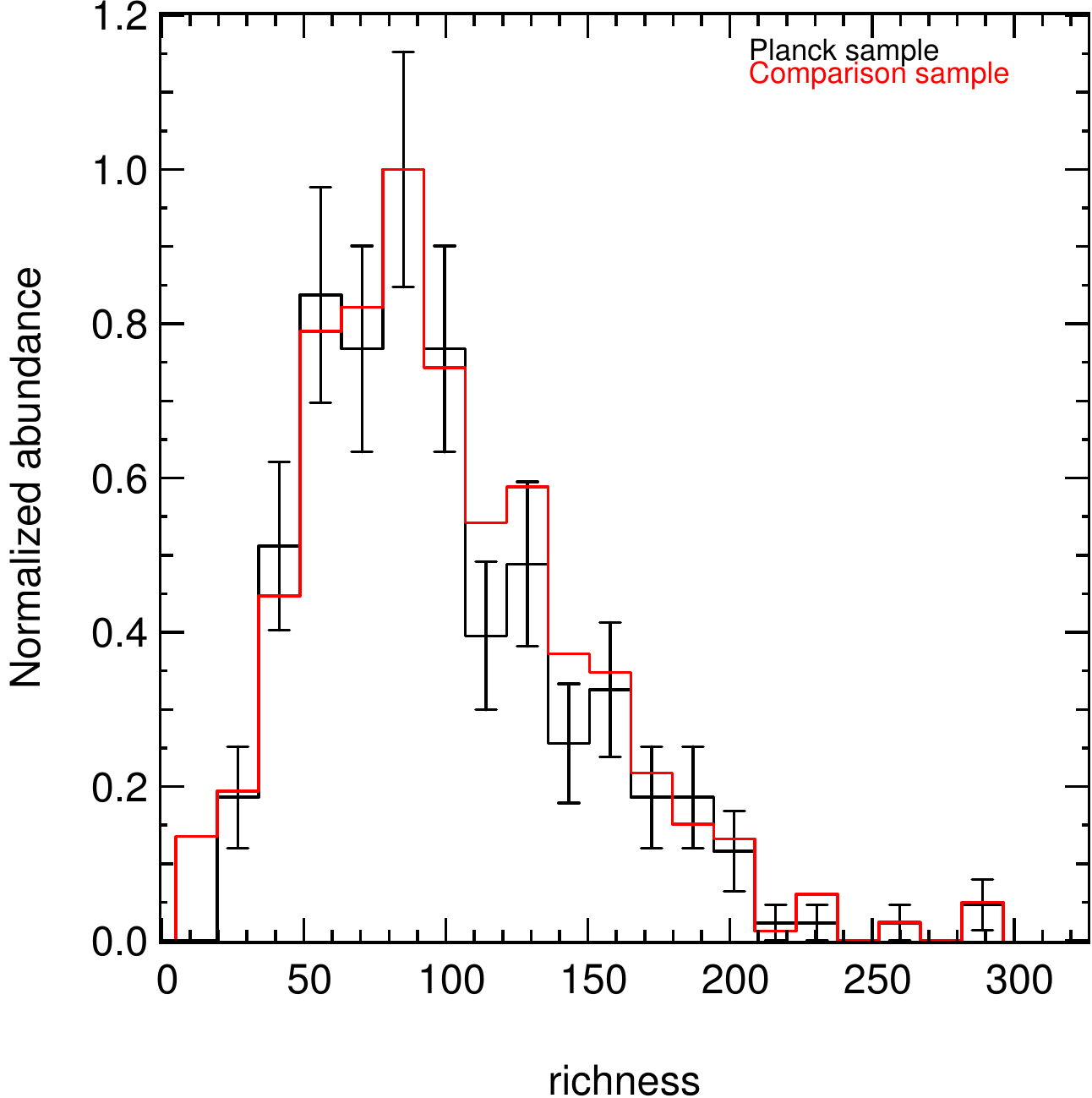}}
  \caption{Redshift (left) and richness distribution (right) of the Planck sample (black) and the comparison sample (red).
  Planck error bars are Poissonian. Comparison sample error bars are not shown but are of comparable size.}
  \label{selection}
\end{figure*}

\begin{figure}
  \centering
  \resizebox{\hsize}{!}{\includegraphics{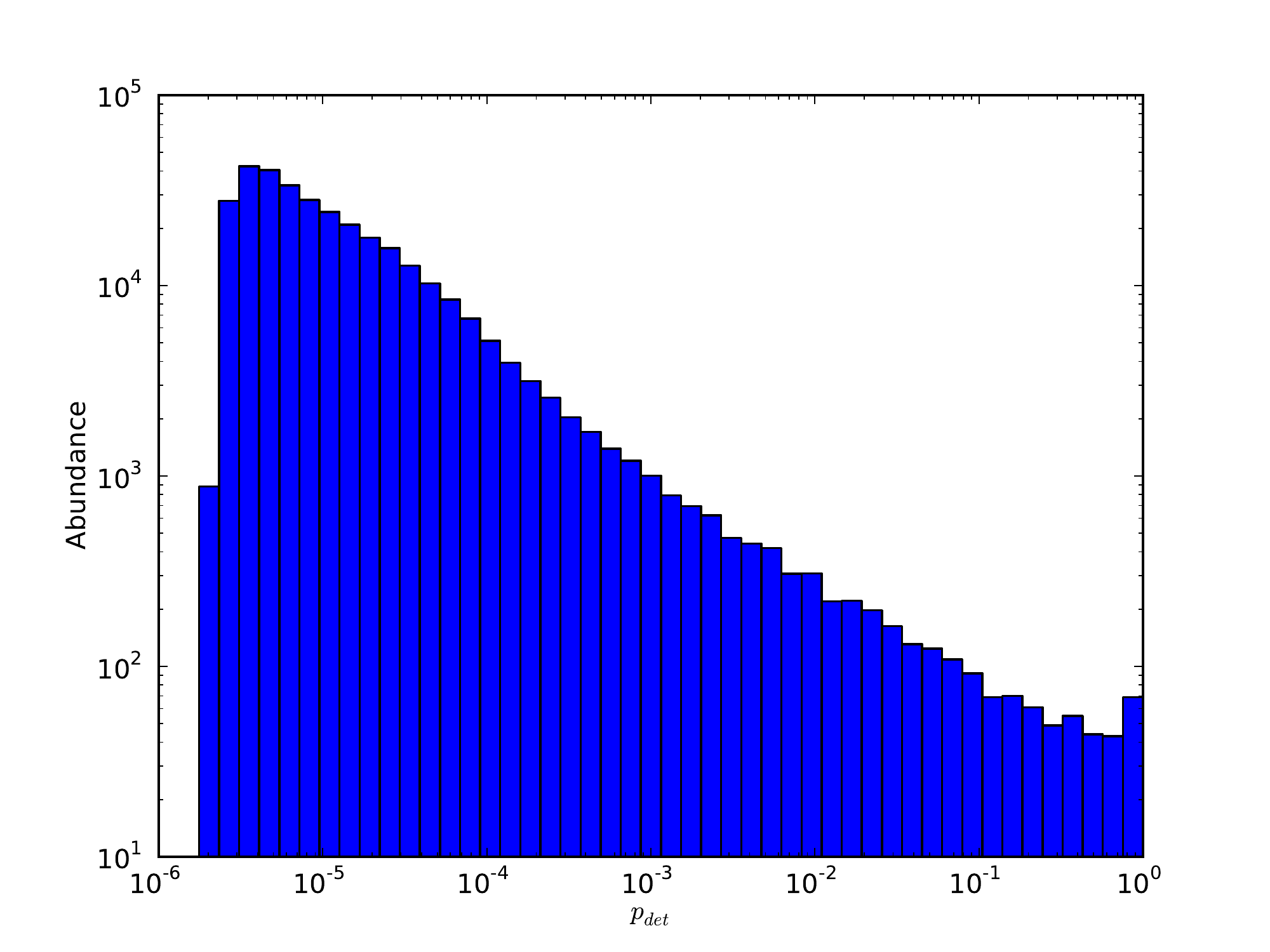}}
  \caption{Log-log histogram of the detection probability.}
  \label{pdet_histo}
\end{figure}

\begin{figure}
  \centering
    \includegraphics[scale=0.41]{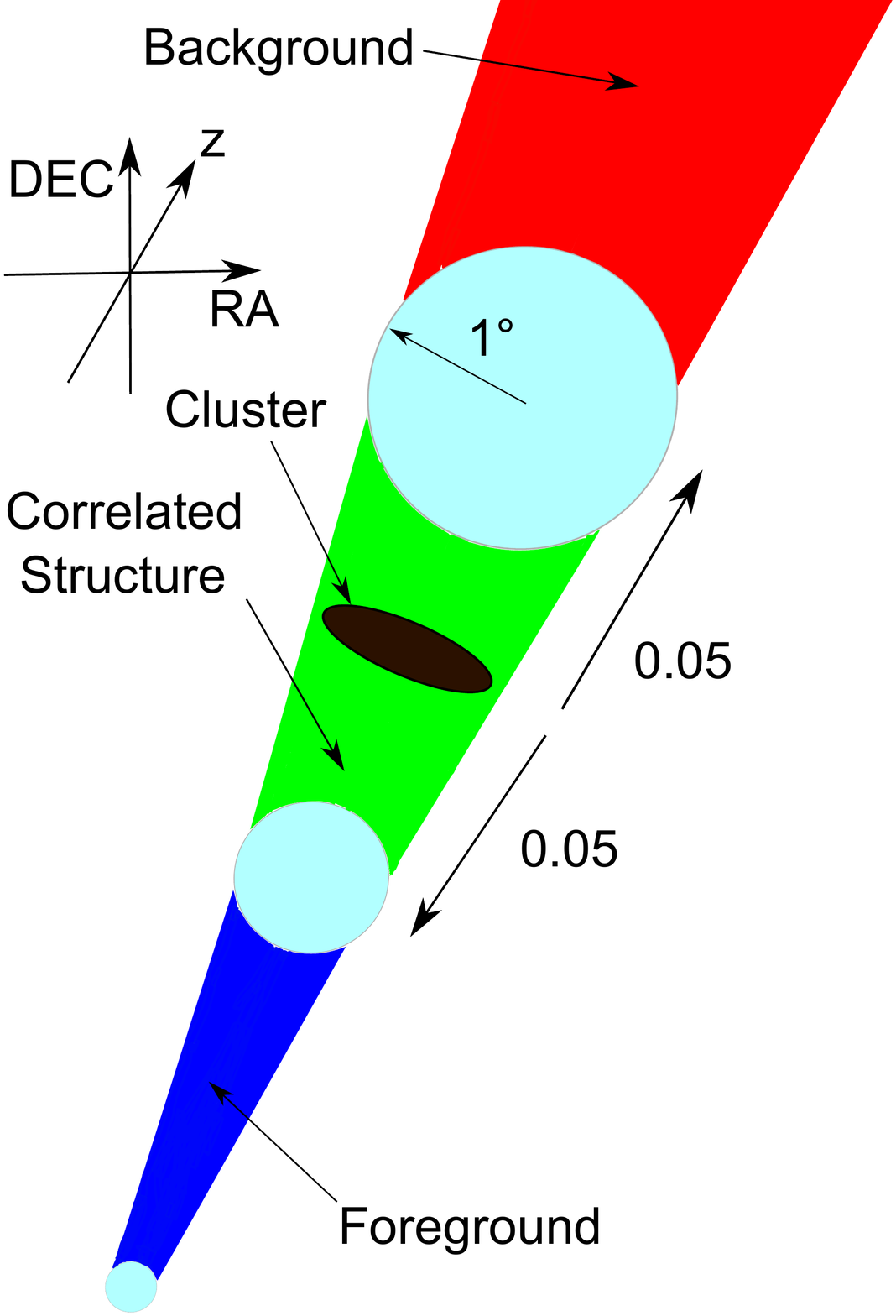}	
  \caption{Sketch of the selection method for the group catalogs. The green, blue and red volumes show the selection for the correlated, foreground
  and background samples respectively. The total z-depth of the green volume is
  0.1, with the cluster in the center. The angular radius for all volumes is 1$^\circ$ (see cyan circles).}
  \label{z_cylinder}
\end{figure}

\section{Methods}
\label{sect3}

\subsection{Two-point correlation function}

We measure the crowding of clusters and groups with the angular two-point correlation function,
which traces the amplitude of cluster/group clustering as a function of their separation.
The angular correlation function $w(\theta)$ is defined as the excess probability over a
random, uncorrelated distribution of finding two objects separated by an angle
$\theta$. The probability of finding two objects in two infinitesimal solid angle
elements $\delta \Omega_1$ and $\delta \Omega_2$ separated by angle $\theta$
then reads:

\begin{equation}
\delta P = n_1 n_2 \left ( 1 + w(\theta) \right ) \delta \Omega_1 \delta \Omega_2 ,
\end{equation}
\noindent
with $n_1$ and $n_2$ being the mean cluster/group densities in both samples.


\noindent
A multitude of different estimators exist for calculating the two-point correlation function from data catalogs. \citet{2000ApJ...535L..13K}, who
compared the nine most important of these estimators in terms of the cumulative probability of returning a value within a certain tolerance
of the real correlation, show that the Landy-Szalay estimator \citep[hereafter LS,][]{1993ApJ...412...64L} performs best according to their criteria. Hence 
we adopt this estimator, which reads:
\begin{equation}
  \hat{w}_{LS}(\theta)=\frac{DD-2DR+RR}{RR} ,
\end{equation}

or, in case of a cross-correlation between two different samples:
\begin{equation}
\label{eq_LS}
  \hat{w}_{LS}(\theta)=\frac{D_1D_2-D_1R_2-D_2R_1+R_1R_2}{R_1R_2} ,
\end{equation}

where $DD$, $DR$ and $RR$ stand for the data-data, data-random and random-random pair counts, respectively. All pair counts in eq. \ref{eq_LS}
are normalized to the total number of data pairs in the respective samples. Random points account for geometrical effects
like survey boundaries and masks in the survey area. We do not want the random points to correct for environment-based detection effects, since this
is the effect we want to measure, so we are using random points where the true detections have been erased.
The pair counts have been computed using the 2d-tree code \textit{Athena} \citep{2014ascl.soft02026K}.

\subsection{Generation of random points for the Planck catalog} 
\label{sect3_rnd}

The LS estimator (eq. \ref{eq_LS}) needs a random catalog for each data catalog, in order to correct for
geometrical effects that could mimic a signal.

There are two effects that might imprint a spatial variation on the \textit{Planck} detection function: the variation in the noise level and the
distance from the galactic disk. In this section we describe our approach 
to generate random points for the \textit{Planck} catalog taking into account the varying noise level. The variation of the detection probability
as a function of distance from the galactic disk is investigated in appendix \ref{appA}.

Since the noise level of the \textit{Planck} observations varies over the SDSS region, we need to test whether the density of SZ detections
has a significant correlation with the noise level that has to be accounted for when generating a random catalog.
We use the \textit{Planck SMICA} map (which comes in \textsc{Healpix} \citep{2005ApJ...622..759G}
coordinates with $N_{\mathrm{side}}$ =2048, which is 50331648 pixels, resolution $\sim$ 1.7$'$), which uses an optimal combination of the nine
frequency bands \citep[]{2013arXiv1303.5072P} to generate a map displaying the weighted average noise of all channels, averaged to 3072 pixels,
to find the noise at the position of each cluster.

We test for correlation of the density of \textit{Planck} detections with the noise quantitatively. In this case we assume the number of Planck
detections per unit area $f_i$ to be a power law of the noise $N$ per redshift bin $i$ with redshift dependent exponent $\alpha_i$:
\begin{equation}
\label{noise_powerlaw}
 f_i(N)=f_i(1) \cdot N^{\alpha_i} .
\end{equation}

We perform a likelihood analysis over the parameters $f_i(1)$ and $\alpha_i$, by calculating the expected number of clusters in each sky cell
via eq. \ref{noise_powerlaw} and computing the Poisson probability with the actual number of detections in that sky cell.
The power $\alpha_i$ scatters around and is consistent with zero for redshifts $z\leq0.5$. Above this redshift, we find
$\alpha_i \approx 0.8$. 
In conclusion, the noise level has no impact on detections for $z<0.5$. 
We decide to remove all clusters with $z>0.5$ from our catalog, bringing our sample size down to 250 clusters.
Based on these findings, we decide to use uniformly distributed random points for the \textit{Planck} catalog.

We use the Planck survey mask (\textsc{Healpix} $N_{\mathrm{side}}$ =2048) to define the region where to 
generate the points and cut them afterwards to the SDSS footprint. The random points are generated in \textsc{Healpix} coordinates to ensure a uniform
distribution over the sky.

\subsection{Generation of random points for the \textit{RedMaPPer} catalog}

To generate a random point catalog for \textit{RedMaPPer}, we first draw a random position in the sky, and then randomly draw a \textit{RedMaPPer} cluster.  
Given the assigned cluster redshift and richness, we use our cluster model to randomly draw cluster galaxies to create a synthetic cluster. 
We then run \textit{RedMaPPer} at this location, and determine whether the synthetic cluster is detected of not.   
The procedure is repeated 100 times, and we calculate the fraction of times $w$ that the cluster was detected at that location.  
The quantity $w$ is the weight assigned to this random point.

There is one subtlety associated with the above procedure: by random luck, some fraction of our synthetic clusters will overlap with 
real \textit{RedMaPPer} clusters in both location in the sky and redshift.  If one did not remove the galaxies associated with the 
original \textit{RedMaPPer} cluster before placing the synthetic cluster at that location, upon running \textit{RedMaPPer} one will always find
a cluster there (i.e. the original cluster), and one would erroneously conclude $w=1$ irrespective of the details of the synthetic cluster.
Thus, it is critically important to remove the original \textit{RedMaPPer} galaxy clusters from the galaxy catalog prior to drawing our random points.
We erase clusters probabilistically: given a cluster of richness lambda at redshift z, we collect all of its member galaxies, and remove each 
galaxy according to the assigned membership probability, so that a galaxy that is 90\% likely to be a cluster member is removed from 
the galaxy catalog with 90\% probability.

\subsection{Definition of a comparison sample} 
\label{sect3_3}

We want to test whether SZ selected clusters are generally found in a different environment than similar (in terms of redshift and richness)
clusters that are selected for their optical properties.
Thus we need to compare the cluster-group two-point correlation functions that we obtain
for groups in the vicinity of \textit{Planck} selected clusters to an independent sample of optically selected clusters that resembles the selection function
of our main sample in terms of their redshift and richness distribution. To this end, we need to model the \textit{Planck} detection
probability. 
We assume that the probability that a \textit{RedMaPPer} cluster is detected by \textit{Planck} takes the form:
\begin{equation}
  P_{\mathrm{det}}= \frac{1}{2} \left [1 +  \mathrm{erf} \left( \frac{\Lambda-\Lambda_{\mathrm{det}}}{\sqrt{2} \sigma} \right) \right] ,
 \end{equation}
 
 where erf is the error function, $\Lambda_{\mathrm{det}}$ is the richness at which the detection probability is 50\% and $\sigma$ the scatter in richness
 at fixed SZ signal.  $P_{\mathrm{det}}$ states the probability that a cluster of given richness $\Lambda$ is detected by the \textit{Planck} survey, if it was
 inside the survey area. We use the \textit{RedMaPPer} SDDS DR8 catalog, calculate $P_{\mathrm{det}}$ for each cluster and assign it as a weight
 to the cluster itself.
 
 We parameterize the redshift evolution of $\Lambda_{\mathrm{det}}$ and $\sigma$ as:
 \begin{equation}
  \Lambda_{\mathrm{det}}=\alpha_{\Lambda} (1+z)^{\beta_{\Lambda}}
 \end{equation}
 and
 \begin{equation}
  \sigma=\alpha_{\sigma} (1+z)^{\beta_{\sigma}} .
 \end{equation}
 
 To find the optimum values for $\alpha_{\Lambda}$, $\beta_{\Lambda}$, $\alpha_{\sigma}$ and $\beta_{\sigma}$, we perform a likelihood
 analysis in these four parameters:
 \begin{equation}
  \mathrm{ln(L)}= \sum_{i\, \mathrm{Plck}} \mathrm{ln} \left[ \mathrm{P}_{\mathrm{det}}(i) \right ] 
  + \sum_{i\, \mathrm{non\,Plck}} \mathrm{ln} \left [ 1-\mathrm{P}_{\mathrm{det}}(i) \right ] . 
 \end{equation}
 
 Here the first sum is over all \textit{RedMaPPer} clusters that have been detected by Planck and the second sum is over all \textit{RedMaPPer} clusters that have not been
 detected by Planck. Figure \ref{selection} shows the photo-z distribution of the Planck sample (black) compared to the subsample (red) defined by
 the selection algorithm based on detection probability. The data agree in most bins within 1$\sigma$ (of the Poissonian errors) and in all
 bins within 2$\sigma$. To validate the quality of the comparison sample we drew 1000 random subsamples of 250 clusters according to their
 $P_{\mathrm{det}}$ and determined the likelihood of each subsample. Comparing to the likelihood of the original
 \textit{Planck} sample, we obtain a p-value of 0.27, so we consider our comparison sample as reasonable (i.e., 27\% of subsamples have lower
 likelihood than the actual \textit{Planck} sample).
 
 We generate a random catalog for the comparison sample by using the derived values for the four parameters $\alpha_{\Lambda}$, $\beta_{\Lambda}$,
 $\alpha_{\sigma}$ and $\beta_{\sigma}$ and calculate the detection probability for each entry in the \textit{RedMaPPer} random catalog.

\subsection{Theoretical two-point correlation function}
\label{sect3_4} 
 
Our purpose is to compute the cross-correlation between a reference
cluster at given redshift and correlated structures within
a certain redshift range. Note that this differs from computing
the usual angular correlation function between two samples. In
our case, in fact, we restrict our calculation of the cluster-group
two-point correlation function to redshift bins centered around the
reference cluster. The correlated group redshift distribution is thus
dependent on the reference cluster redshift. We then obtain the
total correlation function by summing up all the redshift-binned
contributions according to the cluster redshift distribution.

The numerical tool we use for calculating the theoretical correlation function is \textsc{camb sources} 
\footnote{http://camb.info/sources/}\citep[]{Lewis2007}, which computes the angular power spectrum $C_l$s of the matter density perturbations,
for given input redshift distributions and for different cosmological models. We restrict our calculation to standard flat $\Lambda$CDM cosmology 
($\Omega_m=0.25$, $h=0.7$) 
and the linear regime only. The relation between the cross-spectra and the projected two-point correlation
function is given by
\be
w(\theta)=\sum_{l \geq 0} \left( \frac{2l+1}{4\pi} \right)\, P_l(\cos\theta)\,C_l\ ,
\label{eq:wtheta_cls}
\ee

where $P_l$ are the Legendre polynomials of degree $l$. 
We use a maximum $l=3000$ and $\theta \in [0.01,300]$ arcmin.
We calculate the expected two-point correlation (eq. \ref{eq:wtheta_cls}) for 20 reference cluster redshifts $z_{\mathrm{cl}} \in  [0.05; 0.5]$.
For the reference cluster redshift distribution, we assume a Gaussian distribution centered at the cluster redshift $z_{\mathrm{cl}}$,
with standard deviation equal to the
mean photometric redshift error associated to the cluster redshift
in the Planck catalog, i.e. $\mathcal{N}(z_{\mathrm{cl}}, 0.02)$. For the correlated groups
redshift distribution, we use the observed redshift distribution of
the \textit{RedMaPPer} groups with richness $\lambda > 5$, limited to a range of
$\pm$0.06, centered around $z_{\mathrm{cl}}$. This interval 
is greater than the bin width in the analysis
of the observational data of $\pm 0.05$ (see section \ref{sect4}), in order to account for the errors in photometric redshift ($\sim 0.02$).
The observed correlation is the average of the $w_i(\theta)$ calculated in each redshift bin $i$,
weighted by the cluster and group distributions and respective average biases, normalized by the total number of objects:
\be
w(\theta)_{\mathrm{theory}}= \frac {\sum_{i=1}^{20}  \mathrm{d}N^{\mathrm{c}}_{i} \, \mathrm{d}N^{\mathrm{g}}_{i}\, \bar{b}_i^{\mathrm{c}}\,
\bar{b}_i^{\mathrm{g}}\, w_i(\theta)} {\sum_{i=1}^{20} \mathrm{d}N^{\mathrm{c}}_{i} \, \mathrm{d}N^{\mathrm{g}}_{i}} \ .
\ee
Here $\mathrm{d}N^{\mathrm{c}}_{i}$ and $\mathrm{d}N^{\mathrm{g}}_{i}$ are the counts per redshift bin of clusters and groups respectively.
Furthermore, $\bar{b}_i^{\mathrm{c}}$ and $\bar{b}_i^{\mathrm{g}}$ are the average biases for clusters and groups within the bin $i$. 
We estimate the bias for each cluster in the matched \textit{Planck} catalog and each group in the \textit{RedMaPPer} catalog by using the 
analytic formula of \citet{Tinker2010}. This assumes a fixed mass-richness scaling relation, for which we employ the result of \cite{Rykoff2012}. 
An analog estimate for the foreground/background structures at $\left | z_{\mathrm{cl}} - z_{\mathrm{gr}} \right | >0.05 $ yields a 2pcf consistent with zero
within the statistical errors of our analysis.

\section{Analysis and results}
\label{sect4}

\begin{figure*}
  \centering
  \resizebox{0.8\hsize}{!}{\includegraphics{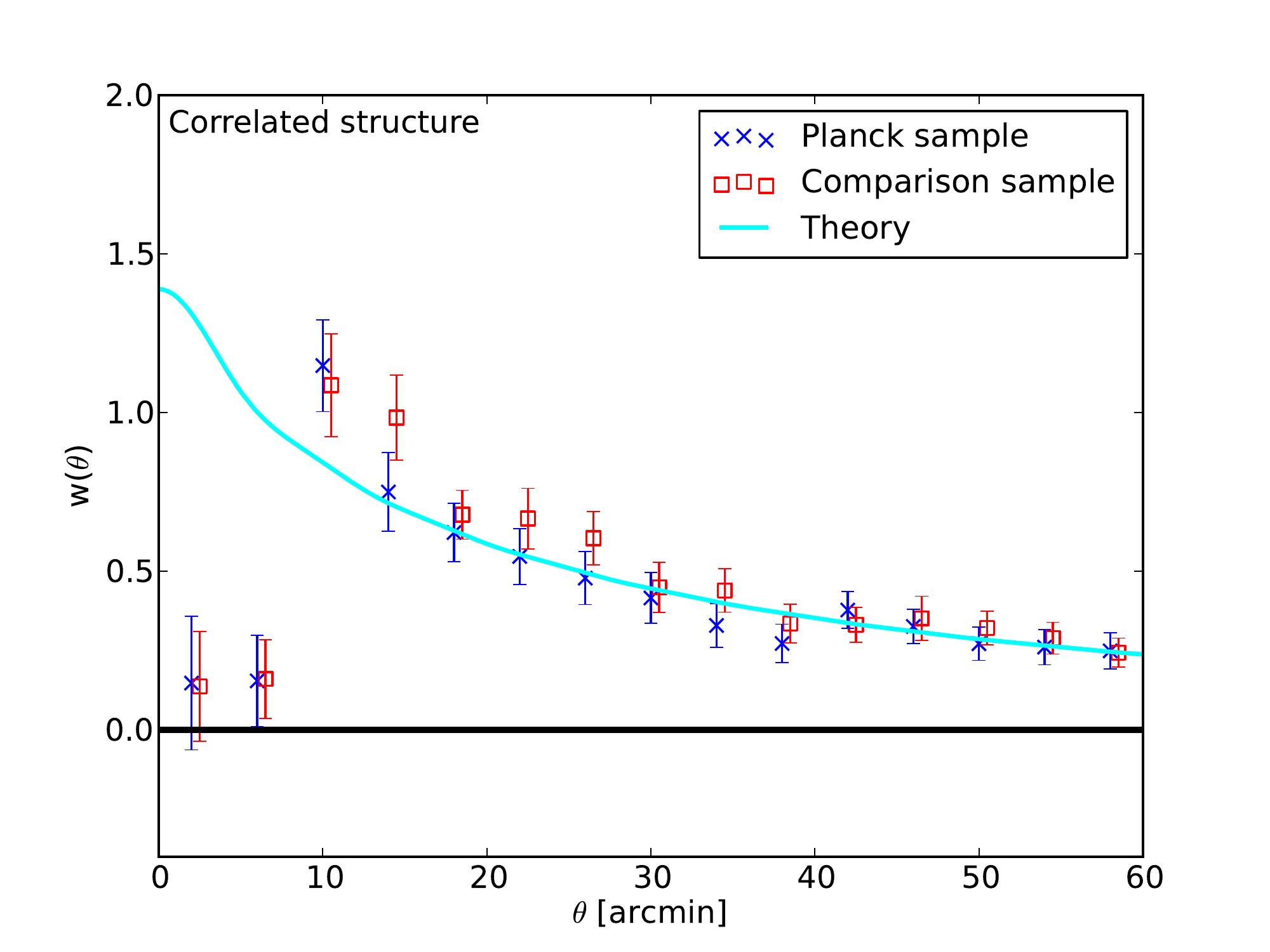}}
  \vspace*{2ex}
  \resizebox{\hsize}{!}{\includegraphics{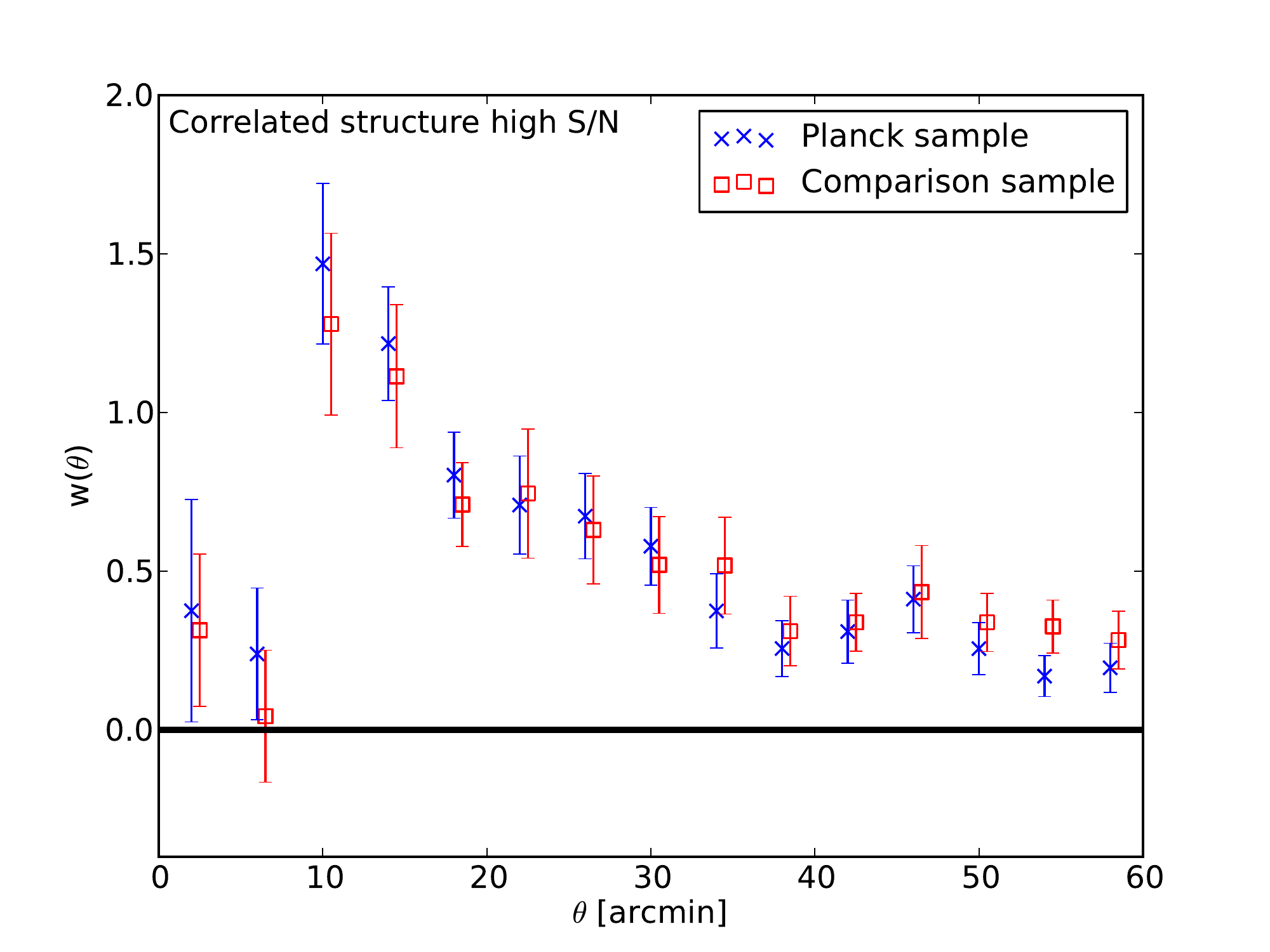}
\hspace*{20mm}
\includegraphics{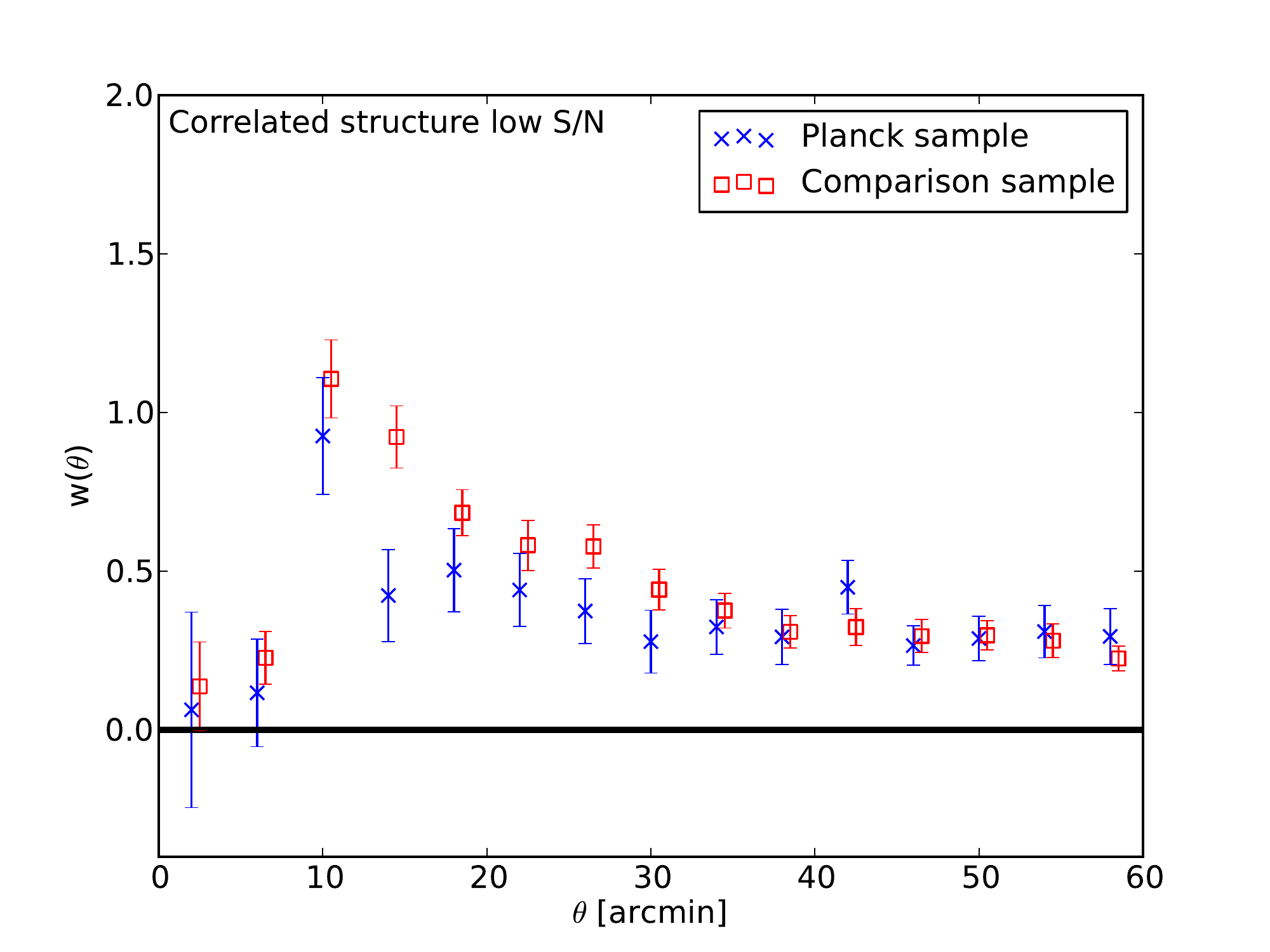}}
  \caption{Two-point correlation function for groups in the vicinity of \textit{Planck} clusters (blue) and
  groups in the vicinity of clusters in our comparison sample (red). In this plot we show the 2pcf for groups with redshift equal to the cluster
  redshift $\pm 0.05$ (correlated structure). The cyan line represents the theoretical prediction. Top: complete sample, bottom left: only clusters with
  S/N \textgreater median, bottom right only clusters with S/N \textless median. In the low S/N case there is a slight underdensity in the
  \textit{Planck} sample in the region between 10$'$ and 20$'$. For interpretations see sections \ref{sect4_2}, \ref{sect4_3} and \ref{sect5}.}
  \label{2pcf_cor}
\end{figure*}


\begin{figure*}
  \centering
  \resizebox{0.8\hsize}{!}{\includegraphics{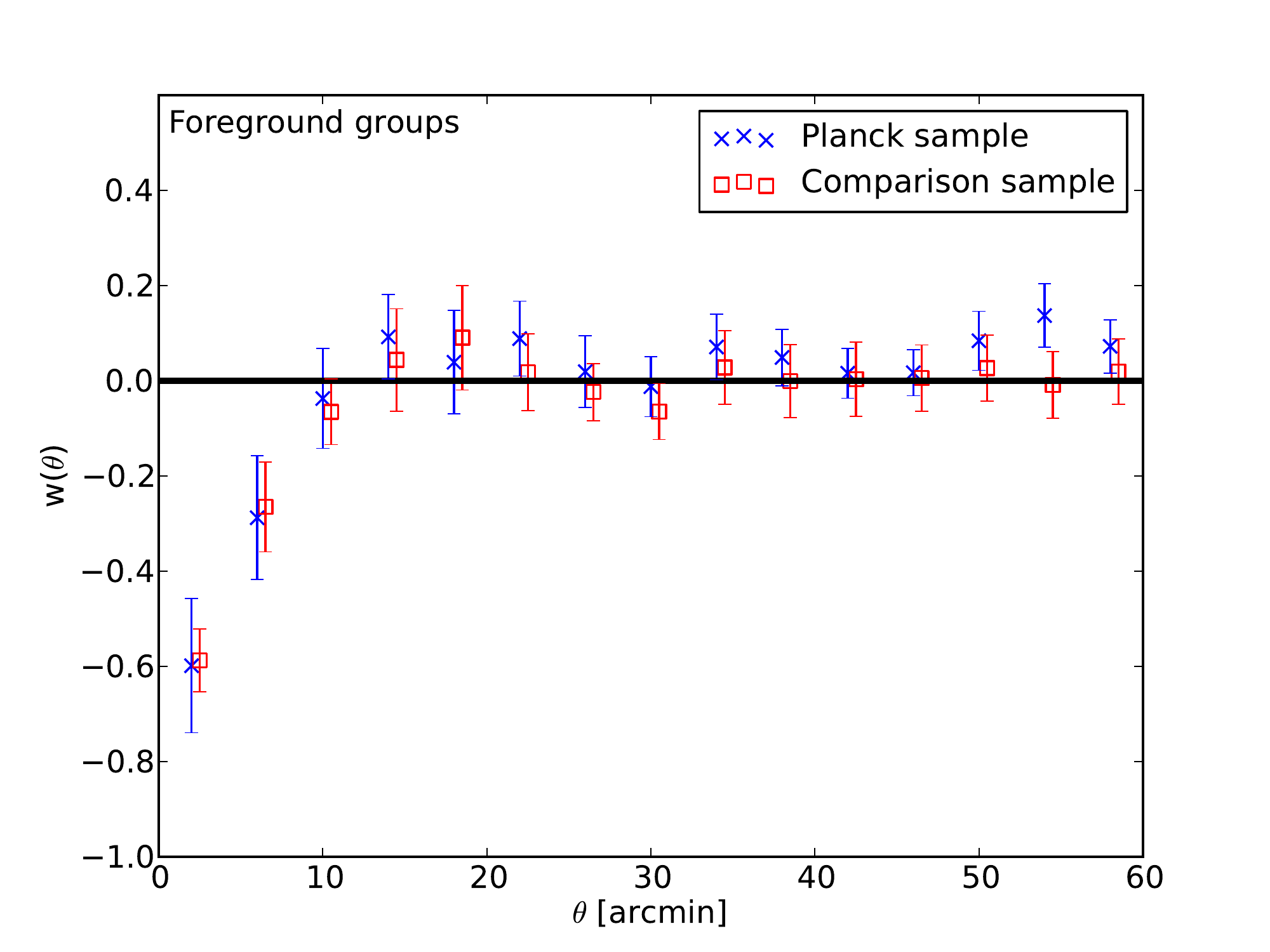}}
  \vspace*{2ex}
  \resizebox{\hsize}{!}{\includegraphics{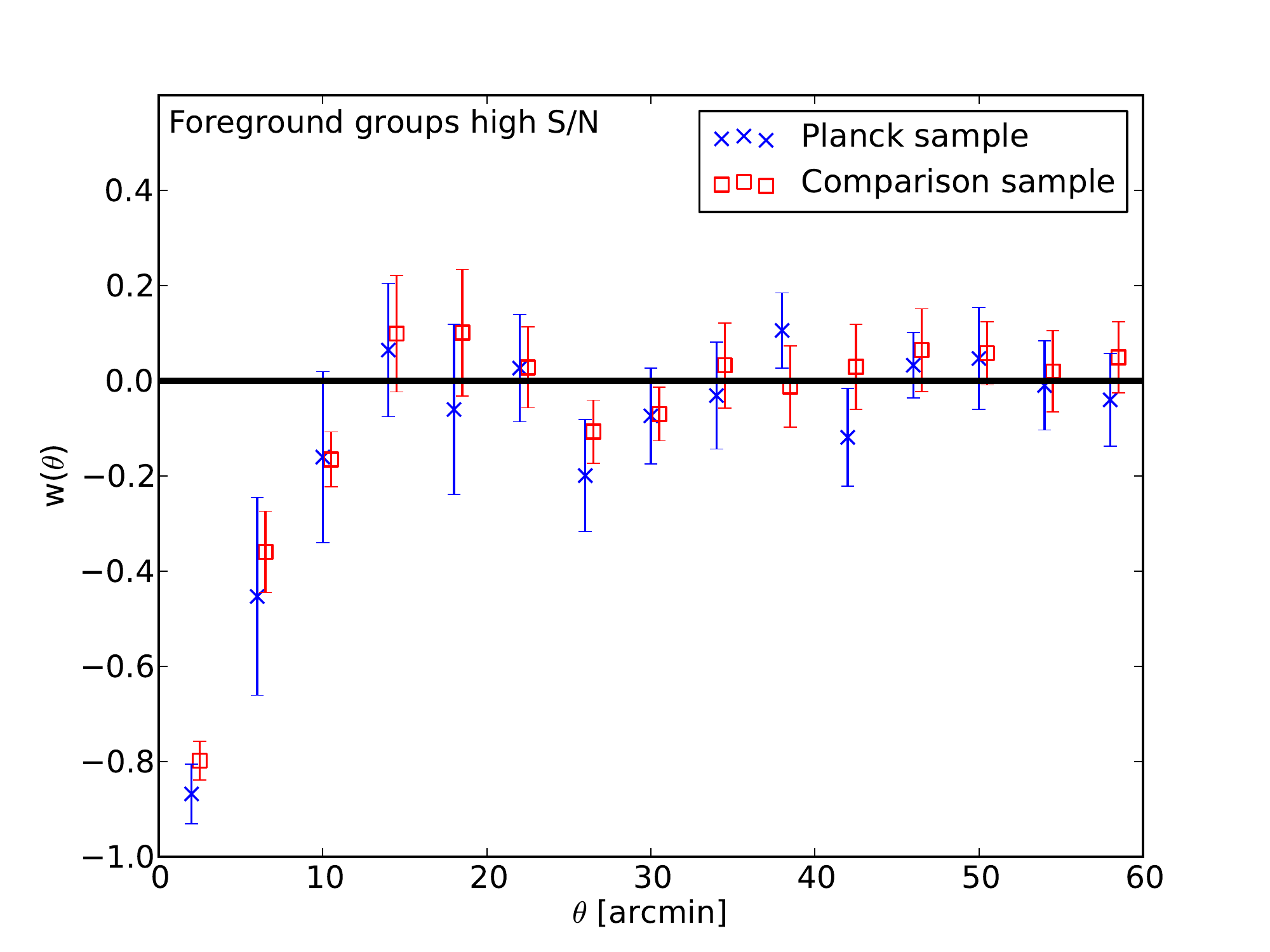}
\hspace*{20mm}
\includegraphics{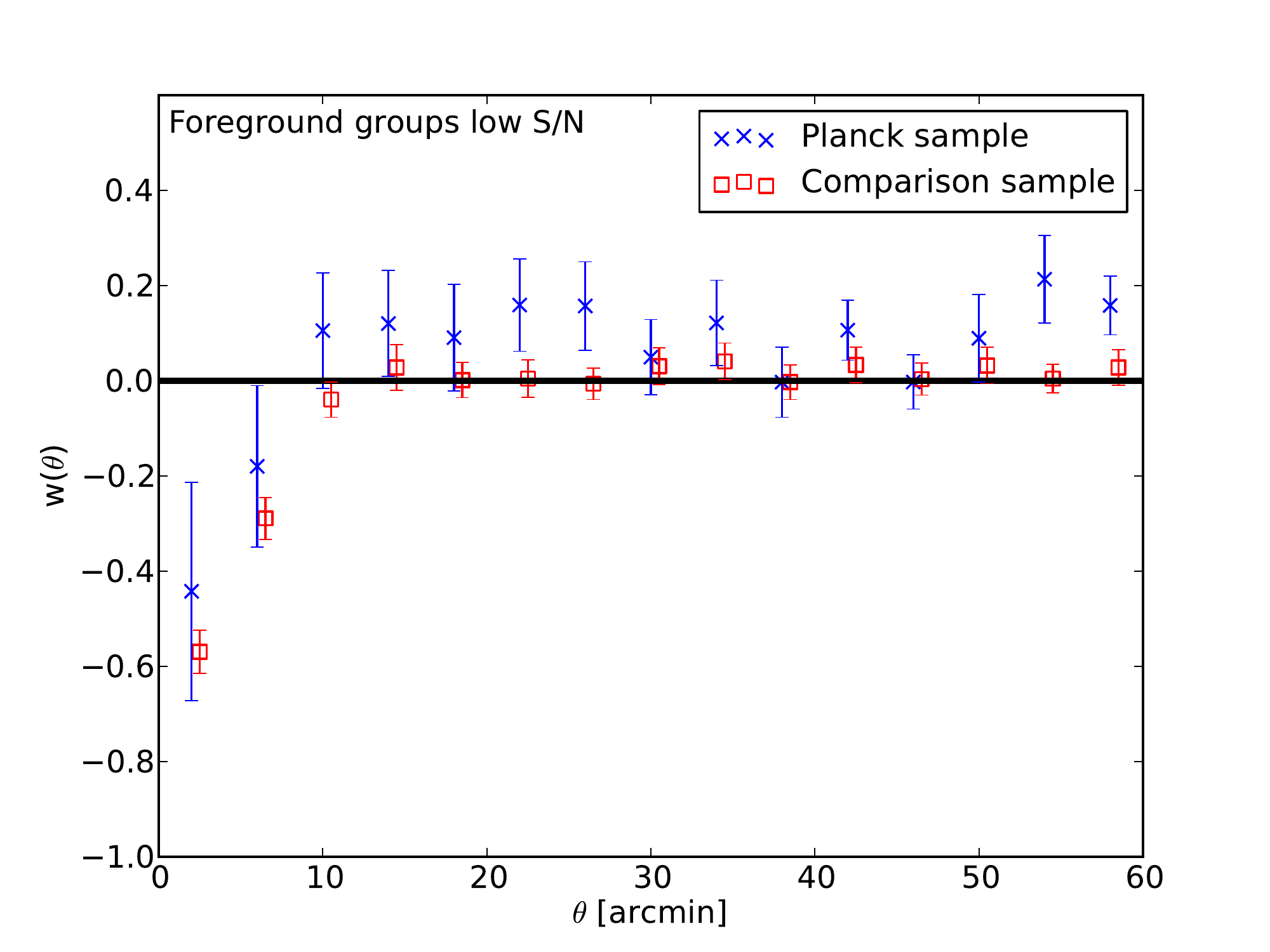}}
  \caption{Same as figure \ref{2pcf_cor}, but for groups with redshift $z_{\mathrm{gr}} < z_{\mathrm{cl}}-0.05$ (foreground structure).
  The two data sets agree well in the complete sample and the high S/N case, while for low S/N a slight overdensity can be observed in the \textit{Planck}
  sample nearly over the complete angular region tested, albeit most data points still agree within the error margins.}
  \label{2pcf_fg}
\end{figure*}

\begin{figure*}
  \centering
  \resizebox{0.8\hsize}{!}{\includegraphics{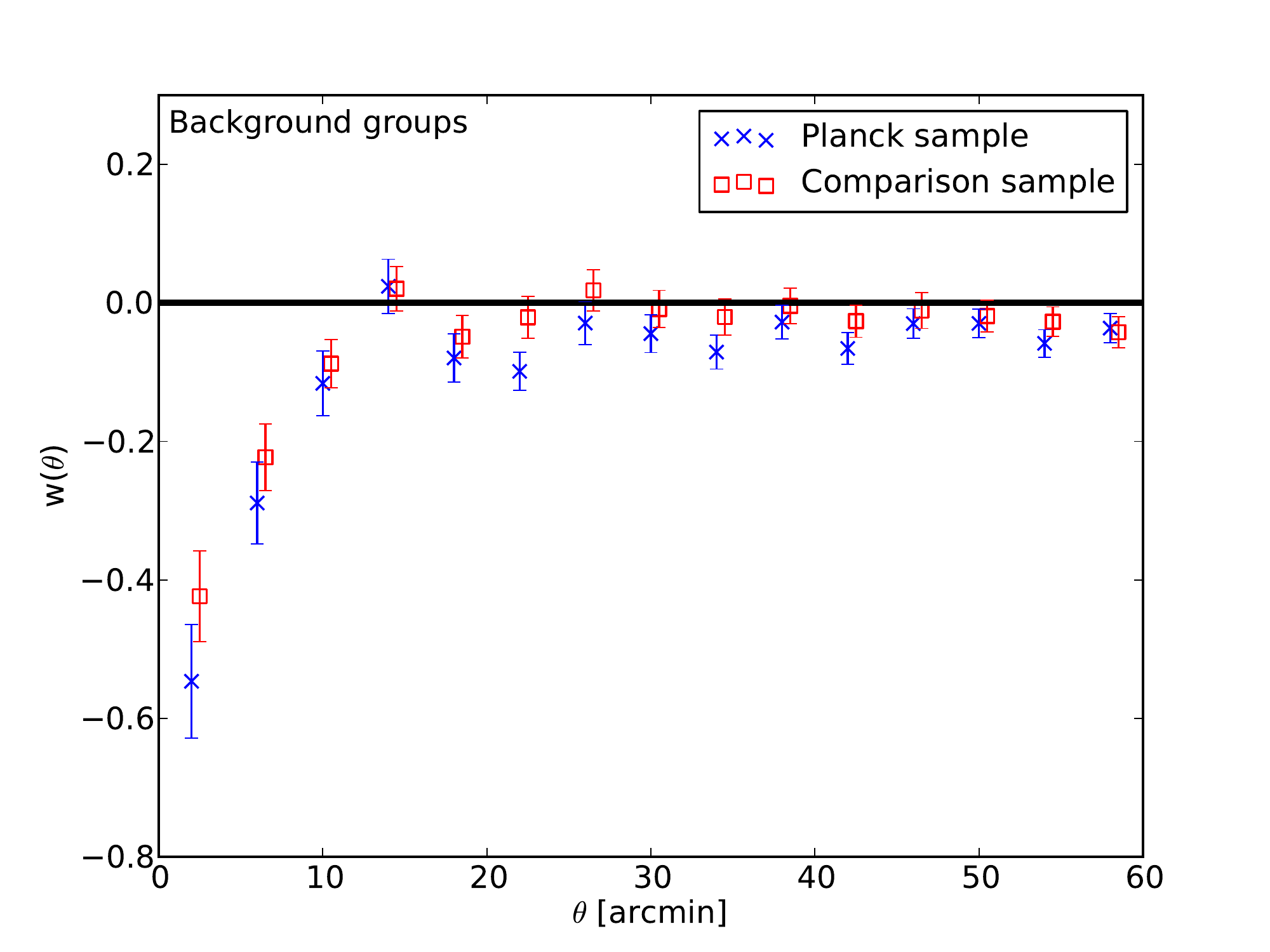}}
  \vspace*{2ex}
  \resizebox{\hsize}{!}{\includegraphics{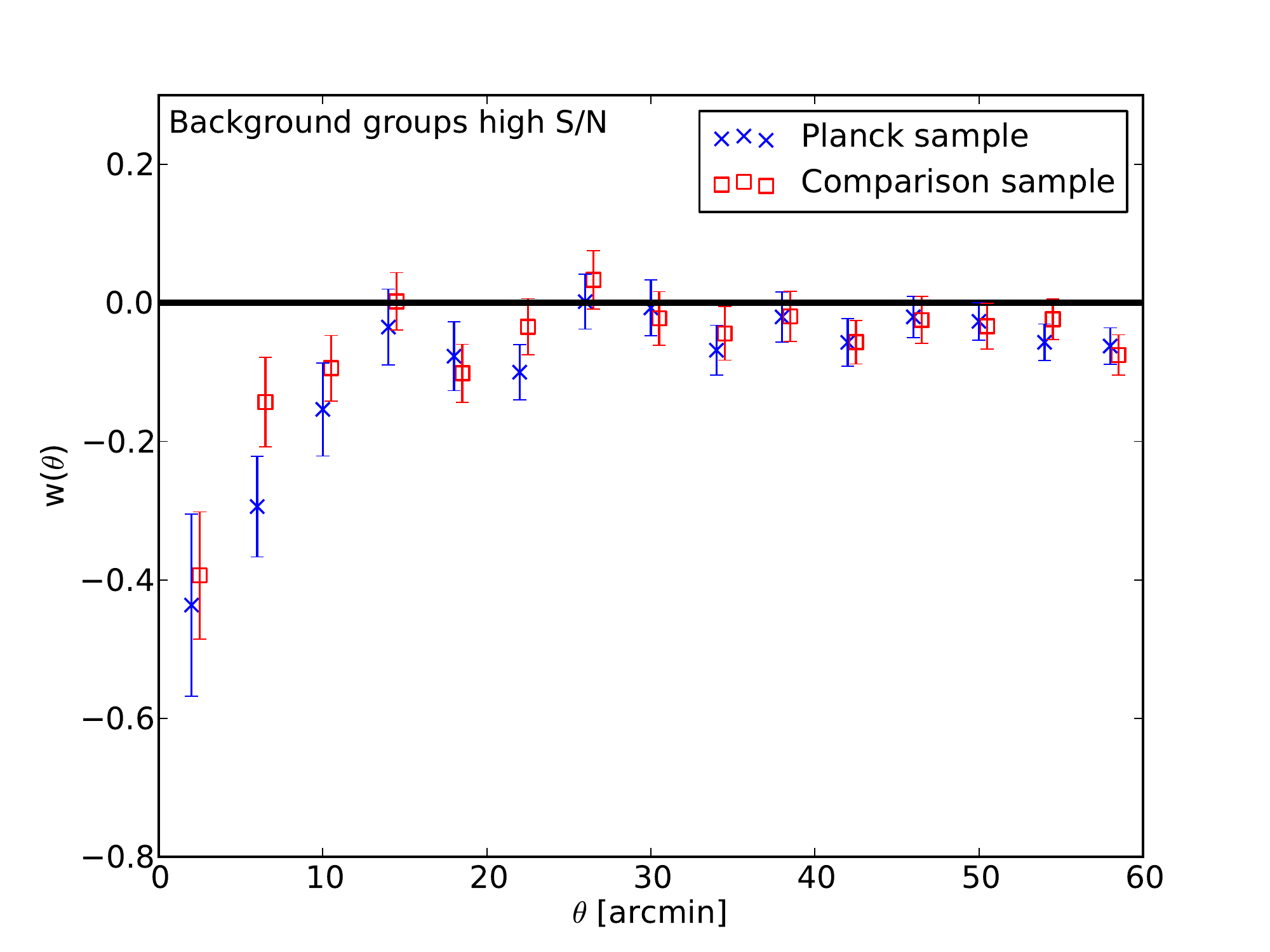}
\hspace*{20mm}
\includegraphics{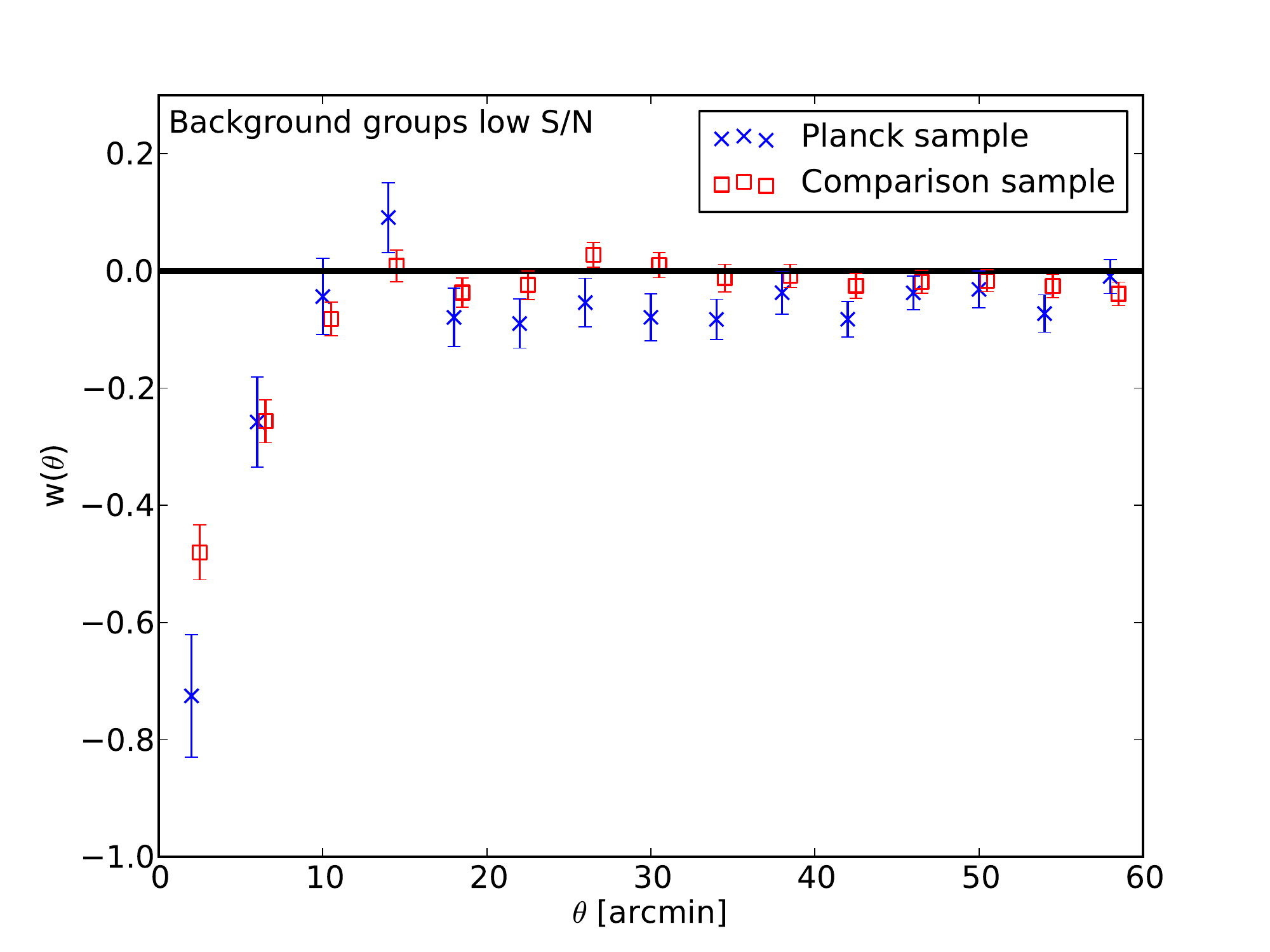}}
  \caption{Same as figure \ref{2pcf_cor}, but for groups with redshift $z_{\mathrm{gr}} > z_{\mathrm{cl}}+0.05$ (background structure).
  We observe a slight underdensity in the \textit{Planck} sample with respect to the comparison sample, which is more severe in the low S/N subsample.}
  \label{2pcf_bg}
\end{figure*}


The relevant quantities we are interested in are the 2pcfs of clusters and groups for correlated structure (groups with similar redshifts as the cluster), 
foreground structure (groups with lower redshift than the cluster) and
background structure (groups with higher redshift than the cluster). 
We compute the angular correlation function, where cluster pairs are subjected to one of three constraints:
\begin{enumerate}
 \item the \textit{RedMaPPer}-\textit{Planck} cluster pair is separated by less than $\left | \Delta z \right |$ \textless 0.05.

 \item the \textit{RedMaPPer}-\textit{Planck} cluster pair is such that $z_{\mathrm{rm}} < z_{\mathrm{pl}} -0.05$.
 \item the \textit{RedMaPPer}-\textit{Planck} cluster pair is such that $z_{\mathrm{rm}} > z_{\mathrm{pl}} +0.05$.
\end{enumerate}

The first set of pairs allows us to test for the environmental impact of physically correlated structures, the second for the impact of foreground
structures and the third for the impact of background structures.
This selection method is displayed graphically in figure \ref{z_cylinder}, with the green volume being the correlated structure, in blue 
the foreground and in red the background.

We draw 100 sets of 250 \textit{Planck} random points, 
assigning them the same redshift distribution as the cluster sample. We perform
the procedure described above on each set, averaging the results.
We proceed analogously for the comparison sample, by drawing
100 samples of unweighted clusters by selecting randomly among all clusters of the comparison sample according to their detection probability. The sample
size is on average 247, the same as the sum over all detection probabilities. The same procedure is performed on the random catalog of the
comparison sample (see section \ref{sect3_3}).

\subsection{Error estimation}

For estimating errors and covariance matrices, we use three different methods:
for the errors of the \textit{Planck} sample with respect to theory, we use a ``replace-one'' implementation of the 
Jackknife resampling method, for the errors of the comparison sample with respect to theory (zero) we use Bootstrap resampling and 
for the errors of the \textit{Planck} sample with respect to the comparison sample we use a ``delete-one'' Jackknife resampling by
drawing 100 different (unweighted) representations out of the complete comparison sample randomly according to the detection probabilities.
The former two will be explained in more detail in the following subsection.

\subsubsection{Errors of the \textit{Planck} sample with respect to theory}

We use a modification of the Jackknife resampling method. In the standard ''delete-one`` Jackknife technique, the survey area is subdivided into a number
of subsamples and the analysis is done 
a number of times equal to the number of subsamples, considering each time all samples except one.
The Jackknife covariance reads:

\begin{equation}
\label{eq:jackcov}
C_{ij} = \frac{m-1}{m}  \sum_{k=1}^m \left( x_{i,k} - \bar{x_i} \right) \left( x_{j,k} - \bar{x_j} \right) ,
\end{equation}

where  $m$ is the number of Jackknife samples,
$x_{i,k}$ is the data value in bin $i$ of sample $k$ and $\bar{x_i}$ is the mean value in bin $i$. 
Since galaxy groups are clustered intrinsically, the errors in neighboring bins may be correlated, so we need to take into
account the full covariance matrix in our analysis.

We define our Jackknife samples to be equal to the data-cylinders in our sub-catalogs. We are using 250 samples containing
exactly one cluster each and all groups in its vicinity.

Since our theoretical prediction is made for the exact redshift distribution of \textit{Planck}, we need to find
the errors with respect to this distribution. A delete-one Jackknife would
introduce a systematic error here, as the redshift distribution of the sample changes when deleting one cluster.
To overcome this problem, we use a modified Jackknife method: in each 
Jackknife sample we leave out one subsample (cluster) and assign a weight of two to another cluster. This cluster is chosen to be the closest
in redshift to the left-out cluster, in order to minimize the effect on the redshift distribution of the sample.
We have to modify equation \ref{eq:jackcov} to account for the changed sample size:

\begin{equation}
\label{eq:jackcovmod}
C_{ij} = \frac{1}{2}  \sum_{k=1}^m \left( x_{i,k} - \bar{x_i} \right) \left( x_{j,k} - \bar{x_j} \right) .
\end{equation}

The validity of the formula has been verified in a Monte-Carlo-simulation.

\subsubsection{Errors of the comparison sample with respect to theory}

In order to perform an error estimation on the comparison sample with a resampling method, we need a multitude of comparison samples. We perform a Bootstrap
resampling on the \textit{RedMaPPer} catalog by drawing 1000 random catalogs with the same number of clusters as in the original catalog. We then count
the number of pairs in angular bins around each cluster weighted with the detection probability and compute the covariance in each angular bin from these
1000 samples. It turns out that the errors estimated by this
method tend to be higher than the errors of the \textit{Planck} sample, since
we did not account for the modified redshift distribution due to the bootstrap
here. To overcome this problem we slightly change the procedure by
bootstrapping sets of 5 groups instead of single groups. The
sets are created by dividing the catalog into 5 subsamples split
by redshift, and selecting one group from each of these subsamples. We sort
the subsamples by weight, so we ensure that each package contains 5 groups
with similar weights and different (equally distributed) redshifts. In this
way the systematic error due to the modified redshift distribution is
minimized.

\subsection{Results}
\label{sect4_2} 
 
In this subsection we present the results of the angular two-point correlation function of galaxy clusters and groups obtained as
described earlier in this section. We analyze $w(\theta)$ in 15 equidistant angular bins
with a width of 4$'$. We compare the results obtained for the \textit{Planck} sample (blue points in figures \ref{2pcf_cor}, \ref{2pcf_fg} and 
\ref{2pcf_bg}) with those for our comparison sample (red points) and with our predictions (cyan line).
A likelihood analysis is presented in subsection \ref{sect4_3}.

We expect that a possible effect is stronger for clusters that are just above the detection threshold S/N of 4.5. Due to this reason we also split
the clusters into a high and low S/N sample. The most useful approach here would be to split the sample at S/N 7, which is the threshold above
which the clusters are included in the \textit{Planck cosmological sample} \citep{2013arXiv1303.5080P}. Unfortunately, in this case the high
S/N sample would contain too few clusters causing the error limits to become too large, so we decided to split the sample at the median S/N 5.4, generating
two equally large subsamples with 125 clusters each.

The top of figure \ref{2pcf_cor} shows $w(\theta)$ for groups at the same redshift as the cluster redshift $\pm 0.05$ (correlated structure).
In the two innermost
angular bins both samples are affected by blending effects and halo exclusion. The latter is the effect of two nearby structures merging into one 
halo, which has not been included in the theoretical prediction. In most bins up to approximately 40$'$ the \textit{Planck} sample shows a slight underdensity
with respect to the comparison sample, albeit the individual data points still agree within the error margins (likelihood analysis shows 
the underdensity is not significant, see table \ref{tab_chi2}). The excess in the third bin with respect
to the predicted curve is potentially due to non-linear structure growth. In case of the split sample we see a better agreement between
the two samples for the high S/N subsample (bottom left plot), while the agreement is worse in the low S/N case (bottom right) where \textit{Planck} clusters
are found in even more underdense background environments.

Figure \ref{2pcf_fg} shows the 2pcf for groups with redshift $z_{\mathrm{gr}} < z_{\mathrm{cl}}-0.05$ (foreground structure). The 
fact that we also observe blending here (in the innermost bins), shows that the detection probability of \textit{RedMaPPer} groups also suffers
from blending effects, i.e. \textit{RedMaPPer} is less likely to detect groups in the vicinity of a rich foreground or background cluster. Besides this effect,
 one can see a slight overdensity in the \textit{Planck} sample at angular scales \textgreater 10$'$, but the errorbars 
suggest that this difference is not significant. The effect is again weaker in the high S/N and stronger in the low S/N subsample.

Figure \ref{2pcf_bg} shows the 2pcf for groups with redshift $z _{\mathrm{gr}}> z_{\mathrm{cl}}+0.05$ (background structure).  Here the 2pcf suffers
from blending on small angular scales, too. A slight underdensity can be seen in the \textit{Planck} sample
with respect to the comparison sample in nearly all angular bins. The individual data points are, however, in agreement within the error margins. 
Also here the observed underdensity appears less severe in the high S/N and more severe in the low S/N case.

\subsection{Likelihood analysis}
\label{sect4_3}

In this subsection we investigate the significance by which the 2pcf in the \textit{Planck} sample differ from the 
comparison sample and from the theoretical prediction. We perform a generalized $\chi^2$ analysis that takes into account the full covariance matrix,
since as mentioned before we expect the errors in neighboring bins to be correlated positively due to the clustering of groups.
The generalized $\chi^2$ reads:
\begin{equation}
 \chi_{\mathrm{gen}}^2 = \bm{\delta}^T \cdot C_{ij}^{-1} \cdot \bm{\delta} ,
\end{equation}

where $C_{ij}^{-1}$ is the inverse covariance matrix and $\delta$ is the residual vector, containing the difference between measured and expected values
(where measured values correspond to the \textit{Planck} 2pcf and expected values correspond to either the comparison sample or predicted values)
in angular bins. For the foreground and background sample we compare the results with zero, since the theoretical predictions in these cases
are several orders of magnitude lower than our measurement uncertainty.

\begin{table*}
\caption{P-values for the different samples for \textit{Planck} with respect to the comparison sample
and to the theoretical prediction and for the comparison with respect to the theoretical prediction.}
  \label{tab_chi2}
  \centering
 \begin{tabular}{lcccccc}
\hline
Sample & & All & High S/N & Low S/N & High $\lambda$ & Low $\lambda$ \\
\hline
Correlated & Plck-Comp & 0.805 & 0.555 & 0.433 \\
 & Plck-Theo & 0.901 &  &  \\
\hline
& Plck-Comp & 0.28 & 0.98 & 0.39 \\
Foreground & Plck-Zero & 0.72 & 0.64 & 0.34 \\
& Comp-Zero & 0.34 & 0.18 & 1.0\\
\hline
& Plck-Comp & 0.48 & 0.70 & 0.37 & 0.89 & 0.73 \\
Background & Plck-Zero & 0.0060 & 0.051 & 0.097 & 0.18 & 0.010 \\
& Comp-Zero & 0.16 & 0.023 & 0.64 & &  \\
\hline
 \end{tabular}
\end{table*}

\begin{table*}
\caption{Best fit values and 1-$\sigma$ intervals for the foreground and background samples, for \textit{Planck} and comparison sample. For the background
sample, \textit{Planck} is not consistent with zero within $4 \sigma$ while the comparison sample is consistent with zero within $1.3 \sigma$.}
  \label{tab_conf}
  \centering
 \begin{tabular}{lcccccc}
\hline
Sample & & All & High S/N & Low S/N & High $\lambda$ & Low $\lambda$ \\
\hline
Foreground & Plck-Zero & $0.040 \pm 0.027$ & $-0.014 \pm 0.041$ & $0.071 \pm 0.036$ \\
& Comp-Zero & $-0.00039 \pm 0.021$ & $0.0052 \pm 0.033$ & $0.016 \pm 0.027$ & & \\
\hline
Background & Plck-Zero & $-0.049 \pm 0.012$ & $-0.047 \pm 0.015$ & $-0.046 \pm 0.016$ & $-0.044 \pm 0.016$ & $-0.058 \pm 0.016$ \\
& Comp-Zero & $-0.02 \pm 0.016$ & $-0.037 \pm 0.023$ & $-0.017 \pm 0.022$ & &  \\
\hline
\end{tabular}
\end{table*}

Table \ref{tab_chi2} gives the p-values for all our three different data samples for \textit{Planck}
with respect to the comparison sample, \textit{Planck} compared to theory and the comparison sample with respect to the theoretical prediction.
The four innermost angular bins have been ignored in the $\chi^2$ calculation, since the data in these bins apparently
suffer from halo exclusion and blending effects, which have not been considered in our theoretical prediction. Thus, the number of degrees
of freedom is 11.

The p-values with respect to the comparison sample are typically quite high (the lowest one being 0.28 for the foreground sample), so the null-hypothesis,
which states that the two samples are similar, cannot be rejected. The p-values are generally slightly lower in the low S/N case which supports our 
assumption that selection effects are predominately observed in the low S/N regime.
Nevertheless, we  cannot confirm a selection bias based on our data sample, since the values of the \textit{Planck} sample and the comparison sample
are in agreement everywhere. 

When comparing the \textit{Planck} data with the theoretical prediction, we find high p-values in the correlated and foreground samples, while we find very
low values in the background sample, which suggests a selection effect related to lower background density.
To support this assumption, we look at the p-value of the comparison sample vs zero (for the background sample) which suggests much better
agreement than the value of the \textit{Planck} sample.
When looking at the splitted sample with respect to S/N, the p-values for the \textit{Planck} sample
are higher than for the complete sample in both cases, which comes from the larger uncertainties in the splitted
sample. The p-values for \textit{Planck} relative to the comparison sample in the high S/N case are in good agreement, yet both only marginally agree
with zero, which we assume comes from cosmic
variance. In the low S/N sample however, the agreement of \textit{Planck} with zero is significantly worse than for the comparison sample.
We conclude that the background underdensity for \textit{Planck} clusters is a function of S/N and the effect becomes stronger for low S/N
detections.

We also split the group sample in two subsamples at richness 12 (high $\lambda$ and low $\lambda$ in table \ref{tab_chi2}), but we found no significant 
differences in these two subsamples.

Table \ref{tab_conf} gives the best-fitting constant values for the 2pcf and the corresponding 1$\sigma$ errors. The first four angular bins which suffer from 
blending have been ignored in this fit. We see that the background correlation
is not consistent with zero for \textit{Planck} with more than 4$\sigma$, while the comparison sample is consistent with zero within $1.25 \sigma$,
which can still be due to statistical fluctuations. Since we detect a background underdensity of -0.049 with a significance of $\sim 4 \sigma$ with respect
to zero but the comparison sample also differs from zero with a value of -0.02 at $\sim 1.25 \sigma$, we conclude that statistical fluctuations in the
particular regions used (cosmic variance), likely also contribute to the observed defect of \textit{Planck} background groups, but are no sufficient
explanation of the full observed effect. On the other hand, one could imagine that \textit{RedMaPPer} detections are biased in the vicinity
of massive clusters due to the correlated structure that surrounds them out to large radii, which might affect the detection of groups due to the blending effect,
as discussed in section \ref{sect4_2}.

When looking at the foreground sample, the slight overdensity one might expect from figure \ref{2pcf_fg} is not significant, with a p-value of 0.72.

\subsection{2pcf for \textit{Planck} and LRGs}
\label{sect_4_4}

We want to verify our results by comparing them to an independent sample of background sources. We replace the \textit{RedMaPPer} group catalog
with the \textit{CMASS} catalog of luminous red galaxies (LRG) with spectroscopic redshifts 
\citep[]{2011AJ....142...72E,2013AJ....145...10D,2014MNRAS.441...24A}. As clusters and groups
tend to feature mostly red galaxies, we expect the LRGs to show a similar clustering behavior. Furthermore, if the origin of the underdensity we
observed is truly the presence of radio sources, which tend to cluster at high redshifts, we expect to see the same effect in \textit{CMASS} galaxies.

\begin{figure*}
  \centering
  \resizebox{0.8\hsize}{!}{\includegraphics{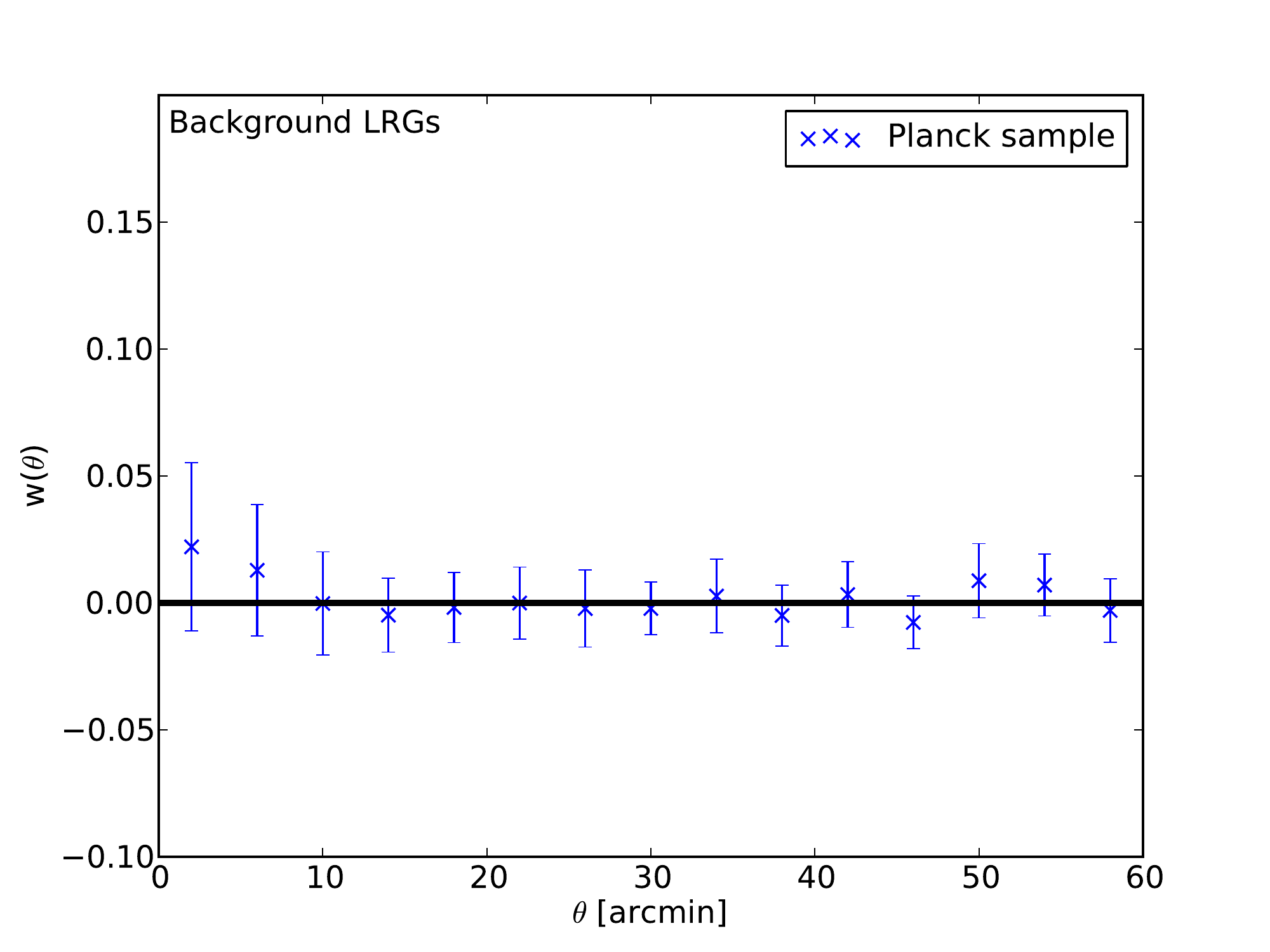}}
  \caption{Same as figure \ref{2pcf_bg}, but for CMASS LRGs with redshift $z_{\mathrm{LRG}} > z_{\mathrm{cl}}+0.05$ (background structure).
  We do not observe the underdensity that we observed for the \textit{RedMaPPer} group background in figure \ref{2pcf_cor}. 
  The small errorbars come from the much larger number of objects due to the non-overlapping redshift distributions 
  of \textit{Planck} and CMASS (all LRGs are background to all clusters with respect to redshift alone).}
  \label{2pcf_lrg_bg}
\end{figure*}

When looking at figure \ref{2pcf_lrg_bg}, we cannot confirm the underdensity that we found for background
\textit{RedMaPPer} groups. If the physical effect has a z-dependence, the fact that the redshift
distributions of the \textit{RedMaPPer} groups and the \textit{CMASS LRGs} differ largely might be responsible for the observed effect.
As we are using uniformly distributed random points for the 
\textit{Planck} clusters, we do not account for a potential position dependence of the selection function.
In particular, a spatial variation of the redshift dependence of the \textit{Planck} detection probability could possibly mimic such a selection effect. We investigated the most likely version of this possibility in Appendix \ref{appA}, 
although more complex dependencies might exist.

\subsection{Implications for SZ and lensing masses}

The potential underdensity in the background of \textit{Planck} detected clusters is nearly constant in all angular bins except the ones that are
affected by blending. Hence it is straightforward to model the effect as the best fitting constant 2pcf in these bins. We use this value of 
-0.049 for further approximations. We estimate the defect SZ
signal caused by this effect in the average beam size of the \textit{Planck} channels that are involved in the cluster detection.
We use the median redshift of the 250 \textit{Planck} clusters in our sample (0.23) and calculate the mean
SZ signal of all \textit{RedMaPPer} groups with redshift higher than that value +0.05 (as our background selection), using the scaling relation
from \citet{2013arXiv1303.5080P}. Analogously
we compute the mean SZ signal of the \textit{Planck} clusters themselves. With the number of background groups with respect
to the previously mentioned \textit{Planck} median redshift inside the average beam size of the involved channels and the average
background underdensity we can calculate the $Y_{SZ}$-defect caused by a background underdensity of -0.049. We find a number of
of $1.7 \times 10^{-4}$ relative to the mean signal of the cluster. This is due to the self-similar slope of the $Y_{SZ}$-MOR.
Based on this low number, we conclude that there will be no implications on cluster masses derived from their SZ-signal. 
For the same reason we conclude the effect of the background underdensity on X-ray measurements to be negligible as well.

We estimate the effect of the potential background underdensity on the convergence $\kappa$, which is the quantity that determines the magnification in
gravitational lensing. As an example, we are using a cluster with a mass of $M=3 \times 10^{14} M_{\sun}$, at the \textit{Planck} median
redshift $0.23$ and sources at redshift of $1.0$, assuming that 5\% of the matter between the cluster redshift and the source redshift is missing.
In a radius of 15' around such a cluster, the relative $\kappa$ deficit would then be 
$\sim 80\%$ of the mean $\kappa$ of the cluster. Due to the mass-sheet degeneracy \citep[]{1995A&A...294..411S,1995A&A...297..287S,1997A&A...318..687S},
this large $\kappa$ defect might have a much smaller effect on shear measurements. 
It will, however, have a non-negligible effect on magnification measurements.
Magnification increases the surface area of observed objects with constant surface brightness, leading to higher total
brightness (lower magnitude). On the one hand, the consequence is a higher (observed) galaxy density, as faint background galaxies (just below
the detection limit) might by detected as their brightness increases. On the other hand, the increased surface area also increases the
separation between the magnified objects, leading to a lower (observed) galaxy density, counteracting the first effect. For steep
luminosity functions the first effect is stronger (which is generally the case for blue galaxies), while for flat luminosity functions the second
effect dominates (red galaxies). As a consequence, we expect a negative 2pcf of red background galaxies around clusters caused by this effect.
A potential background underdensity
would counteract this effect, causing a slightly less negative 2pcf at small angular scales and a slightly positive 2pcf at intermediate
angular scales. We estimate the amplitude of 
this effect for a typical \textit{Planck} cluster using equation 10 from \citet{ 2011ApJ...729..127U}. We get a result in the order of
$w(\theta) \lesssim 10^{-2}$ for $\theta=10'$ for the effect caused by the magnification of the cluster itself, while the counteracting effect caused
by the underdensity is in the order of $w(\theta) \approx 10^{-3}$, both of which is too small to be detected in our measurement.

\section{Conclusions}
\label{sect5}

Our main scientific goal was to investigate possible selection effects on SZ selected clusters based on their 
group environment and estimate implications of such an effect on SZ, X-ray and lensing
mass estimates. 

We summarize our results as follows:
\begin{enumerate}

\item We do not find an overall selection effect due to correlated or foreground structure. 
\item We find a potential underdensity of galaxy groups in the background
of \textit{Planck} clusters which manifests in an average 2pcf in an angular range \textless 40$'$ of $-0.049$ with a
significance of $\sim 4 \sigma$. However, we cannot confirm this effect when replacing \textit{RedMaPPer}
groups with CMASS LRGs in our analysis. 
\item This effect grows stronger for low S/N detections and vanishes for high S/N
detections. We find no dependence of the effect on the richness of the groups.
\item We consider three possible explanations for this effect:
\begin{itemize}
\item  An erroneous background estimation in overdense
environments might lead to a lower detection probability of low signal clusters in 
these regions.
The details and relative importance of these effects is likely dependent on the instrumental and survey design and the object detection algorithm.
On the other hand, the fact that \textit{Planck} detections combine six bands makes this explanation rather unlikely.
\item RLAGNs, which tend to cluster at high redshifts \citep{2010MNRAS.407.1078D,1989MNRAS.240..129Y}, contribute to the radio
signal in regions where the background density is high and could suppress low S/N detections.
\item The \textit{Planck} selection function is responsible for this effect.

A spatial variation of the \textit{Planck} selection function that correlates with the spatial
variation of the \textit{RedMaPPer} selection function could mimic the observed background group underdensity. Due to lack of access to the \textit{Planck} selection function we we are not able to
test this at this point. On the other hand, we do get the same results if we
split our sample by distance to the galactic disk, as shown in Appendix \ref{appA}.

\end{itemize}  

\item This potential selection effect has a vanishing impact on SZ and X-ray mass estimates. The implications on lensing mass
estimates are however much larger with an estimated relative $\kappa$ deficit of order unity. 
\end{enumerate}

In the latter context, it is interesting to note that 
\citet{2014MNRAS.442.1507G} found a discrepancy from the self-similar slope ($\beta=5/3$) in the $Y_{SZ}$-mass scaling relation for low S/N \textit{Planck}
clusters, with a slope of $0.76 \pm 0.20$. \citet{2014arXiv1407.7869S} found a slope of the $Y_{SZ}$-mass scaling relation of $1.22 \pm 0.24$ using all
\textit{Planck} clusters detected by the \textit{MMF3} algorithm, and a slope of $1.40 \pm 0.31$ when using only the cosmological subsample 
(S/N \textgreater 7). They made an additional analysis forcing the intrinsic scatter to zero, obtaining even lower results for the slopes,
$0.95 \pm 0.10$ for the full and $1.09 \pm 0.17$ for the cosmological samples. The background
underdensity in \textit{Planck} clusters that we find in this work potentially explains their findings, since that could cause a low-biased lensing
mass, depending on the S/N ratio of the SZ signal, resulting in a shallower slope of the scaling relation. The fact that \citet{2014arXiv1407.7869S}
find a slightly steeper slope in the cosmological sample, supports the assumption of the S/N dependence of this effect. 
\Citet{2014MNRAS.443.1973V}, who compared cluster masses from the \textit{Planck} catalog with weak lensing masses from the
\textit{Weighing the Giants} project, found evidence for a mass dependence in the calibration ratio between the \textit{Planck} mass $M_{Planck}$ and 
the weak lensing mass $M_{\mathrm{wl}}$ which takes the form $M_{Planck} \propto M_{\mathrm{wl}}^{0.68^{+0.15}_{-0.11}}$.
A possible explanation for their findings might be low-biased weak-lensing masses for low-mass clusters, caused by a background
underdensity that dominates at low S/N, as we hypothesize it in this work.

\section*{Acknowledgements}
The authors thank Martin Kilbinger for discussions concerning the \textit{Athena} tree code.
We also acknowledge Ben Hoyle for giving useful instructions on using the \textit{Healpix} software.
Furthermore we thank Francesco Montesano for helpful discussions about methods of error estimation.
We thank Tommaso Giannantonio for giving invaluable advice for the generation of \textit{Planck} random
points. We acknowledge the advice given by Jim Bartlett and Jean-Baptiste Melin concerning the \textit{Planck}
selection function.

\bibliographystyle{mn2e}
\bibliography{References.bib}

\appendix
\section{Splitted sample with respect to galactic distance}
\label{appA}

As mentioned in section \ref{sect_4_4}, we found a discrepancy in our results as we observe an underdensity in the background of
\textit{Planck} selections for \textit{RedMaPPer} groups but not for CMASS LRGs. In order to investigate the cause of this difference we splitted
the sample at the median absolute galactic latitude to find out whether the \textit{Planck} selection function depends on the distance
to the galactic disk, as it could be caused for example by galactic foreground emission. Our uniformly distributed set of \textit{Planck}
random points would not account for such an effect.

 \begin{table*}
\caption{Same as table \ref{tab_chi2}, this time for the splitted samples with respect to distance to the galactic disk and redshift.}
  \label{tab_chi2_ref}
  \centering
 \begin{tabular}{lcccc}
\hline
Sample &  High abs(lat) & Low abs(lat) & High z & Low z \\
\hline
 Plck-Zero & 0.012 & 0.25 & 0.52 & 0.018 \\
 Comp-Zero & 0.17 & 0.32 & 0.86 & 0.0057 \\
\hline
\end{tabular}
\end{table*}

\begin{table*}
\caption{Same as table \ref{tab_conf}, this time for the splitted samples with respect to distance to the galactic disk and redshift.}
  \label{tab_conf_ref}
  \centering
 \begin{tabular}{lcccc}
\hline
Sample &  High abs(lat) & Low abs(lat) & High z & Low z \\
\hline
 Plck-zero & $-0.029 \pm 0.023$ & $-0.023 \pm 0.016$ & $-0.037 \pm 0.017$ & $-0.059 \pm 0.017$ \\ 
 Comp-zero & $-0.039 \pm 0.017$ & $-0.0023 \pm 0.015$ & $-0.0079 \pm 0.0075$ & $-0.032 \pm 0.0099$ \\
\hline
\end{tabular}
\end{table*}
 
 \begin{figure*}
   \centering
   \resizebox{\hsize}{!}{\includegraphics{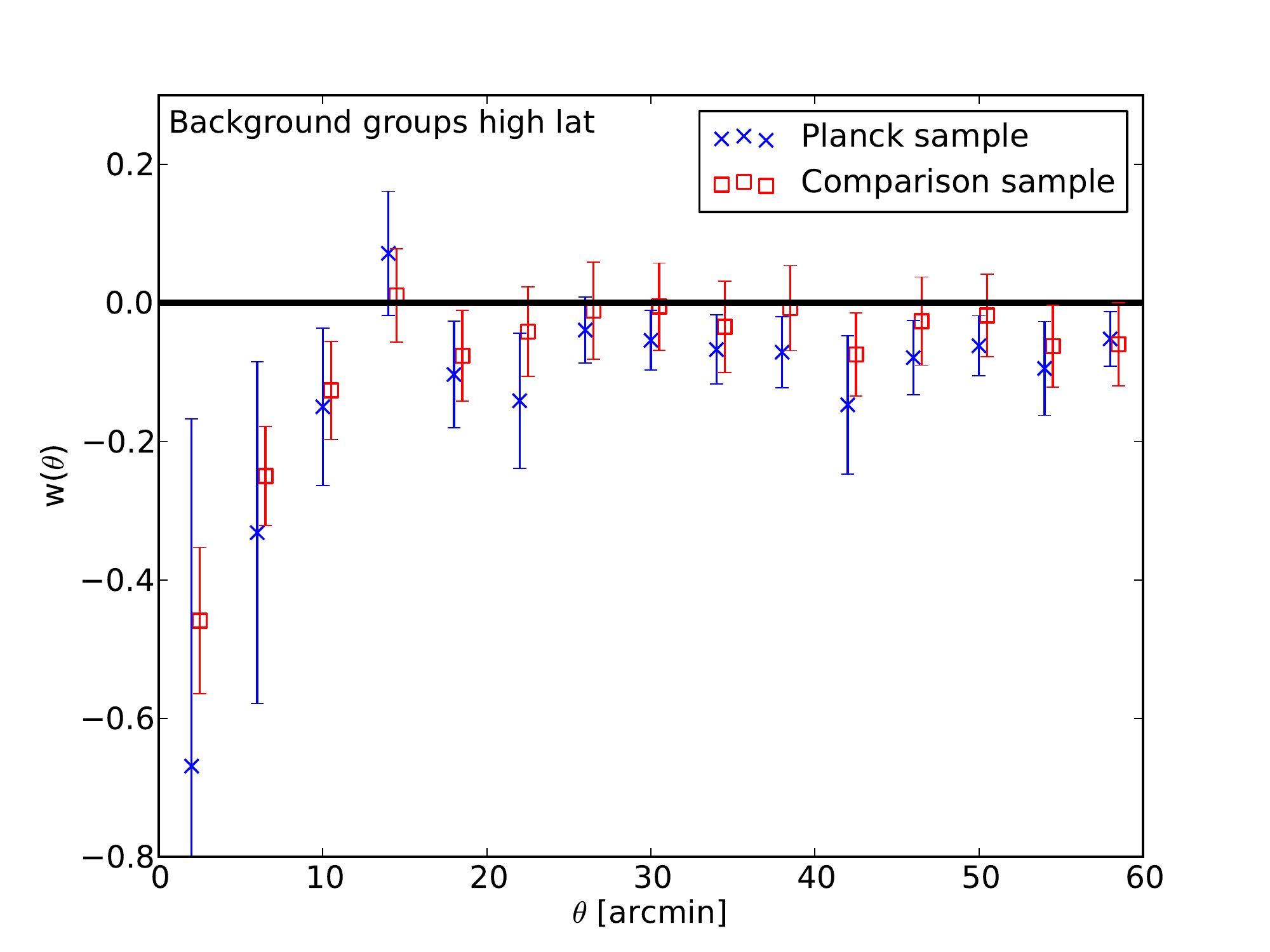}
   \hspace*{20mm}
 \includegraphics{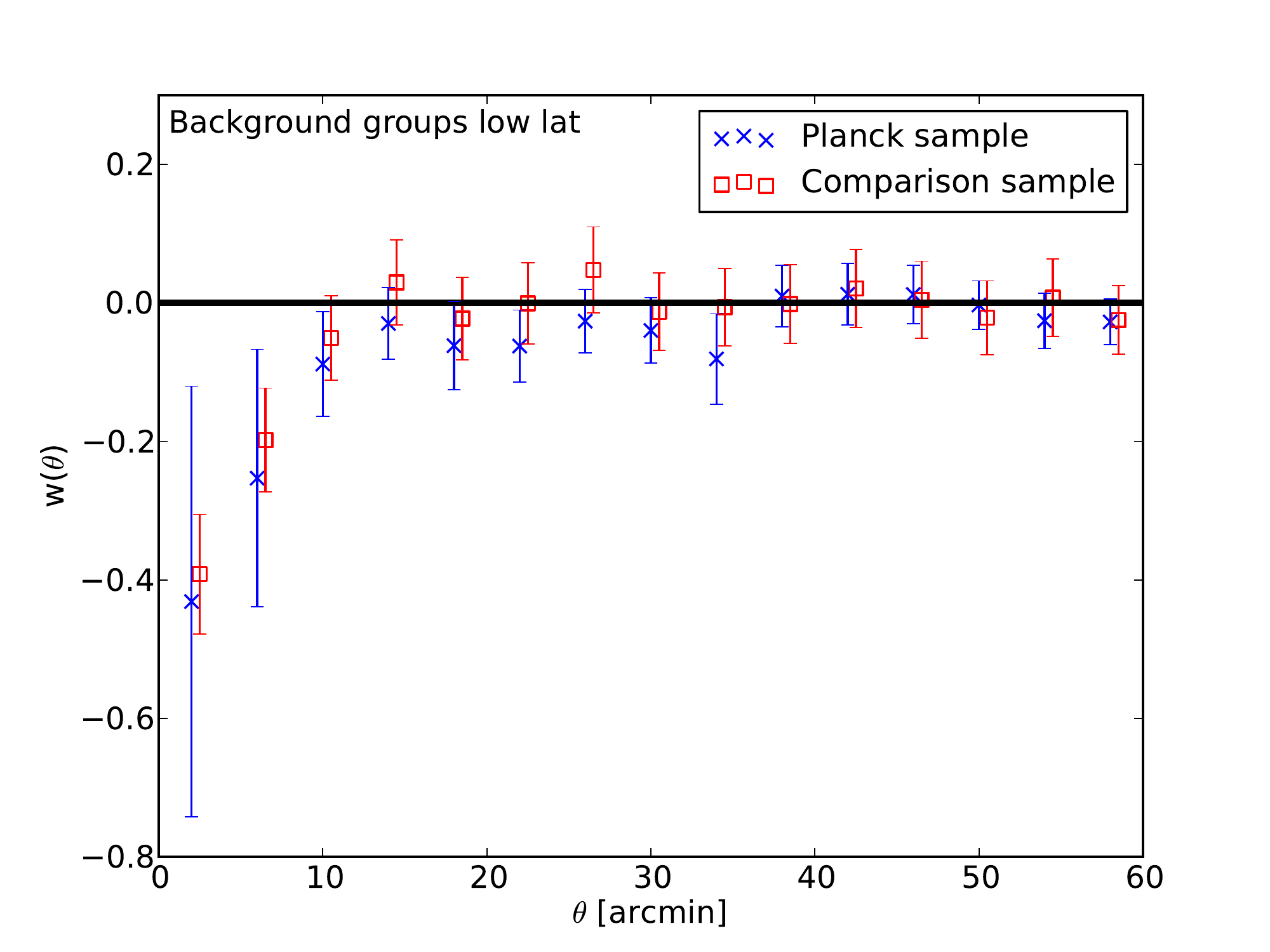}}
   \caption{2pcf for background groups with far from the galactic disk (left) and close to the disk (right). The sample has been splitted
   at the median absolute galactic latitude.}
   \label{2pcf_bg_lat}
 \end{figure*}

 The results of the absolute latitude split are shown in figure \ref{2pcf_bg_lat}, the according
 p-values are found in table \ref{tab_chi2_ref} and the best fitting values and 1-$\sigma$ intervals in  table \ref{tab_conf_ref}.
 The split results in a nearly unchanged result for 2pcf at angular distances up to
 $\sim 30'$. Above that value however, the underdensity vanishes in the low latitude sample while it persists in the high latitude 
 sample.
 
\section{Splitted sample with respect to group redshift}
\label{appB}

As the redshift distributions of \textit{RedMaPPer} and CMASS are largely different, we investigated
the possibility of a redshift dependence of this effect by splitting the \textit{RedMaPPer} group sample at z=0.45. This value has been
chosen to ensure the sample sizes to be approximately equal for the high z and low z sample \textit{after} selecting the background groups.
 
 \begin{figure*}
   \centering
   \resizebox{\hsize}{!}{\includegraphics{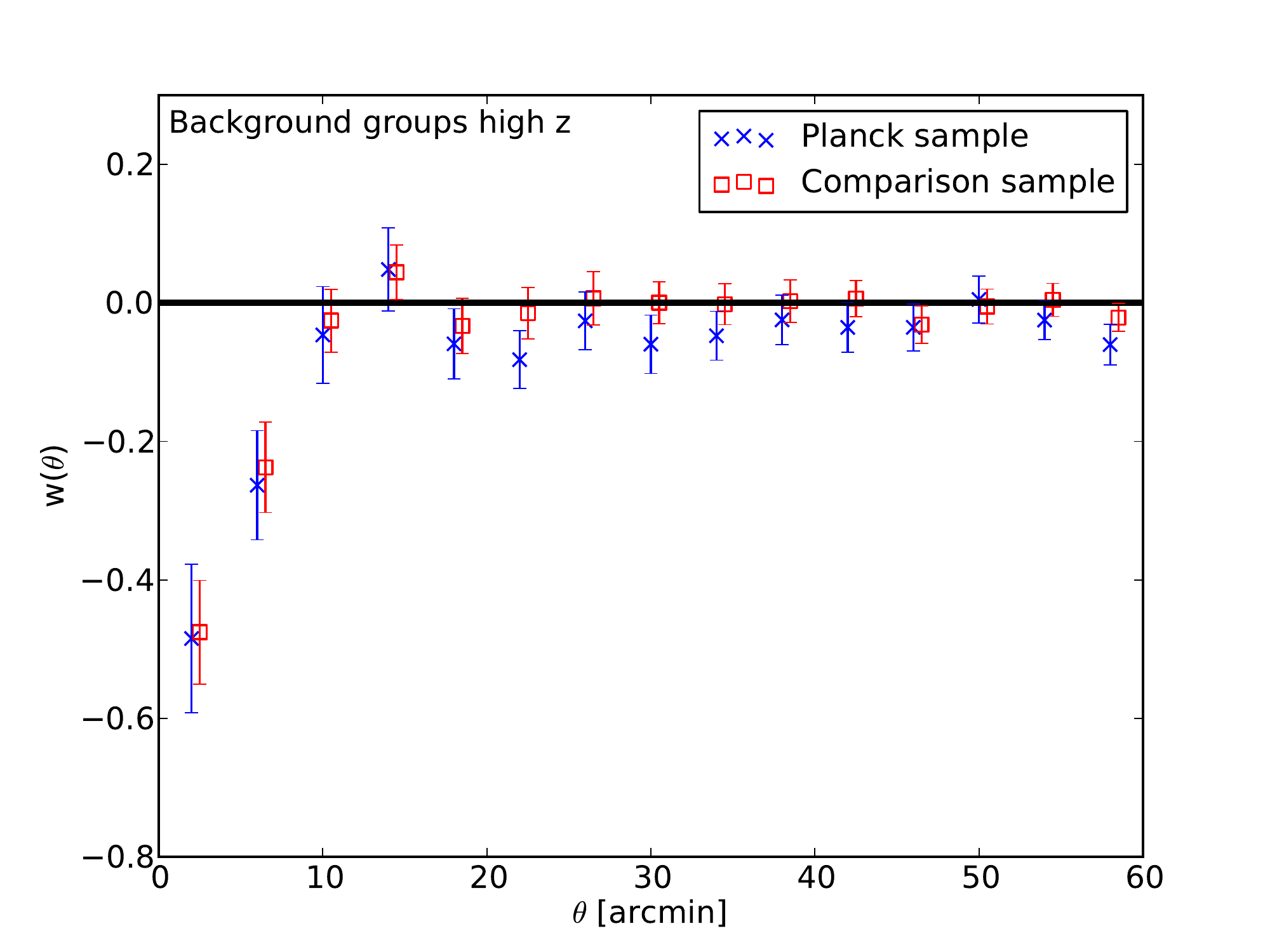}
   \hspace*{20mm}
 \includegraphics{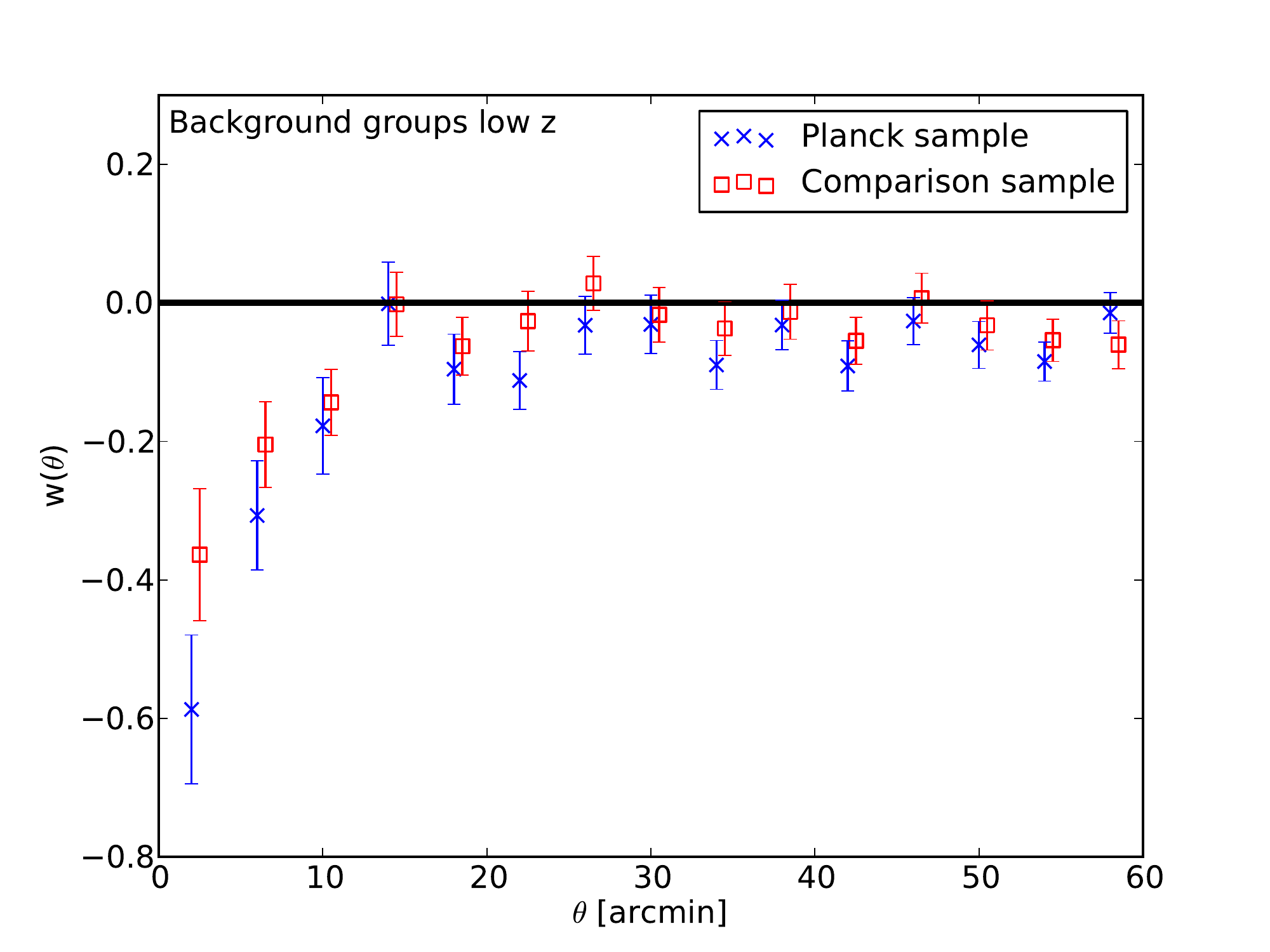}}
   \caption{2pcf for background groups with redshift \textgreater 0.45 (left) and with redshift $\le$ 0.45 (right).}
   \label{2pcf_bg_z}
 \end{figure*}

 The results of the redshift split are shown in figure \ref{2pcf_bg_z},
 while the p-values are shown in table \ref{tab_chi2_ref} and the best fitting values and 1-$\sigma$ intervals in
 table \ref{tab_conf_ref}. 
 The split reveals a slightly more distinct underdensity of the \textit{Planck} sample with respect to zero for low redshift
 background groups, as it might be expected from the null results with the (high redshift) CMASS sample. On the other hand,
 a similar degree of difference between high z and low z background can be observed in the Comparison sample, making the relative difference non-significant.
 
\end{document}